\documentclass[11pt]{article}

\usepackage{amssymb,amsmath}
\usepackage[totalwidth=17cm,totalheight=24cm]{geometry}
\usepackage{graphicx}

\newcommand{\C}{{\mathbb C}}
\newcommand{\R}{{\mathbb R}}
\newcommand{\HH}{{\mathbb H}}

\newcommand{\M}{{P}}
\newcommand{\B}{{B}}
\newcommand{\w}{{\bf w}}
\newcommand{\m}{{\bf m}}
\newcommand{\bt}{{\bf t}}
\newcommand{\e}{{\bf e}}
\newcommand{\f}{{\bf f}}
\newcommand{\W}{{W}}
\newcommand{\E}{{E}}
\newcommand{\F}{{F}}
\newcommand{\cA}{{\mathcal A}}
\newcommand{\h}{{\bf h}}
\newcommand{\A}{{\bf a}}
\newcommand{\id}{{\mathbb I}}
\newcommand{\im}{{\rm i\,}}

\newcommand{\qed}{\hfill\blacksquare}

\newtheorem{proposition}{Proposition}
\newtheorem{corollary}{Corollary}
\newtheorem{definition}{Definition}

\newcommand{\be}{\begin{eqnarray}}
\newcommand{\ee}{\end{eqnarray}}

\begin{document}

\title{6D Interpretation of 3D Gravity}
\author{Yannick Herfray${}^{(1),(2)}$, Kirill Krasnov${}^{(1)}$ and Carlos Scarinci${}^{(1)}$ \\ {}\\
{\small \it ${}^{(1)}$School of Mathematical Sciences, University of Nottingham, NG7 2RD,
UK}
\\
{\small \it ${}^{(2)}$ Laboratoire de Physique, ENS de Lyon,  46 all\'ee d'Italie, F-69364 Lyon Cedex 07, France}}

\date{v2: December 2016}
\maketitle
\begin{abstract} 
We show that 3D gravity, in its pure connection formulation, admits a natural 6D interpretation. The 3D field equations for the connection are equivalent to 6D Hitchin equations for the Chern-Simons 3-form in the total space of the principal bundle over the 3-dimensional base. Turning this construction around one gets an explanation of why the pure connection formulation of 3D gravity exists. More generally, we interpret 3D gravity as the dimensional reduction of the 6D Hitchin theory. To this end, we show that any ${\rm SU}(2)$ invariant closed 3-form in the total space of the principal ${\rm SU}(2)$ bundle can be parametrised by a connection together with a 2-form field on the base. The dimensional reduction of the 6D Hitchin theory then gives rise to 3D gravity coupled to a topological 2-form field. 
\end{abstract}

\section{Introduction}

In this paper we give a new interpretation to gravity in $2+1$ dimensions, by embedding it into a certain theory of a single differential form in 6 dimensions. 

The 6D theory that we consider has been proposed and studied by Hitchin in \cite{Hitchin:2000jd}. This reference introduced a variational principle on the space of 3-forms (in a fixed cohomology class) in 6D manifolds. There are exactly two open ${\rm GL}(6,\R)$ orbits of 3-forms in 6D, one with stabiliser ${\rm SL}(3,\C)$ and the other with stabiliser ${\rm SL}(3,\R)\times{\rm SL}(3,\R)$. Forms belonging to the first orbit are called generic (or stable) of negative type while those in the second orbit are called stable of positive type. In this paper we mainly deal with the more interesting negative type case, but most of our constructions can be easily generalised to the positive case.

As is explained in \cite{Hitchin:2000jd}, a negative type stable 3-form gives rise to an almost complex structure. When the 3-form is taken to be closed and variations are taken inside a fixed cohomology class, the Euler-Lagrange equations following from Hitchin's functional give the integrability conditions for the corresponding almost-complex structure. The 6D manifold thus acquires a complex structure. 

On the other hand, the solutions of 3D gravity (with non-zero cosmological constant $\Lambda$) are constant curvature 3-dimensional metrics. Being constant curvature, every such metric is locally isometric to a homogeneous space. The relevant homogeneous space depends on the sign of $\Lambda$ and the signature. In this paper we will mainly consider the $\Lambda<0$ Euclidean case. We have $\HH_3\equiv{\rm SL}(2,\C)/{\rm SU}(2)$. A related geometric structure is that in the $\Lambda>0$ Lorentzian case when we have ${\rm dS}_3={\rm SL}(2,\C)/{\rm SL}(2,\R)$. Thus, the 3D manifolds in both these cases are locally homogeneous spaces for the group ${\rm SL}(2,\C)$, which is a complex Lie group. 

The relation between the above two constructions stems from the fact that the total space of the frame bundle over either $\HH_3$ or ${\rm dS}_3$ is the ${\rm PSL}(2,\C)$ group manifold and that the standard complex structure on ${\rm PSL}(2,\C)$ arises, in the sense of Hitchin, when one takes the defining 3-form to be the real part of the ${\rm PSL}(2,\C)$-invariant holomorphic 3-form ${\rm Tr}(g^{-1} dg)^3$ with $g\in {\rm PSL}(2,\C)$. This relation, known to experts and to be made explicit below, is the basis for the constructions of this paper.

As we have already stated in the Abstract, what receives a 6D interpretation is the so-called {\it pure connection} formulation of 3D gravity. The only field that appears in the Lagrangian of this formulation is the spin connection, which explains the name. This formulation, first proposed in \cite{Peldan:1991mh} and further developed in Section 3.4 of \cite{Peldan:1993hi}, is not as well-known as it deserves to be. We hope that our work rectifies this, as we show that the pure connection formulation of 3D gravity comes from a natural geometric construction. 

Here is the summary of this construction. Let $\W$ be a connection in the principal ${\rm SU}(2)$ bundle $\M\to M$ over a 3-dimensional base $M$, viewed as a Lie algebra-valued 1-form on the total space $\M$. Consider the 3-form $\Omega=CS(\W)$ on $\M$ given by the Chern-Simons form for $\W$, see (\ref{omega-CS}) below for the definition.
\begin{definition}
A connection $\W$ is called definite if $\Omega=CS(\W)$ on $\M$ is stable (or generic). Connection $\W$ is called negative definite if $\Omega$ is of negative type, i.e. it defines an almost-complex structure on $\M$.
\end{definition}
In section \ref{sec:pure} we will give a different definition of negative definite connections, which is more natural in the 3D context. It will then be seen that the two definitions are actually equivalent. Another remark is that our definition of "definite" connections is inspired by the notion of definite connections on 4-manifolds introduced in \cite{FP}. We also note that the above definition extends immediately also to the 4D case, by taking the Chern-Simons 3-form of an ${\rm SU}(2)$ connection on the principal ${\rm SU}(2)$ bundle over a 4-dimensional base. In this case there is also the notion of generic (or stable) 3-forms in 7D, as well as the notion of sign of such a form, see e.g. \cite{Hitchin:2001rw}. It can be checked that the definition that works with the Chern-Simons 3-form agrees with the definition of definiteness in \cite{FP}. 

Let $J_\W$ be the almost-complex structure defined on $\M$ by $\Omega=CS(\W)$ for a negative definite connection $\W$. In general, an ${\rm SU}(2)$ invariant stable 3-form $\Omega$ on the total space of the principal ${\rm SU}(2)$ bundle defines a connection. This arises because one can declare the image of the vertical vector fields by the almost-complex structure $J$ defined by $\Omega$ to be horizontal. There is also a natural metric on the base which is defined by $J$. This arises by pairing the image of horizontal vector under $J$ using the Killing-Cartan metric along the fibers. Let us denote by $\W(J_\W)$ the connection associated with $J_\W$ and by $E(J_\W)$ a Lie algebra valued basic 1-form that is a frame field for the metric associated with $J_\W$.
\begin{proposition}
Let $\W$ be a negative definite connection, and  let $\E_\W$ be the unique solution of the equation $\F(\W) = \E_\W \wedge \E_\W$. Then the connection $\W(J_\W)$ coincides with $\W$ and $\E_\W$ is a frame for the metric defined by $J_\W$, i.e. $\E_\W=\E(J_\W)$.
\end{proposition}
The frame arising as the solution of this equation will be described explicitly in Section \ref{sec:pure}. Note that the Chern-Simons 3-form $CS(\W)$ is closed because the base is 3-dimensional. Thus, one of the equations of the Hitchin theory $d\Omega=0$ is automatically satisfied. The second Hitchin equation is the statement that $\hat{\Omega}$ is also closed $d\hat{\Omega}=0$, where $\hat{\Omega}$ is obtained from $\Omega$ by acting on all of its 3 indices with $J_\W$. 
\begin{proposition}
For $\Omega=CS(\W)$ the condition $d\hat{\Omega}=0$ is equivalent to $\nabla_\W \E_\W=0$, where $\nabla_\W$ is the covariant exterior derivative with respect to $\W$, i.e. to the statement that the connection $\W$ is torsion-free. 
\end{proposition}
Note that the notion of torsion referred to here is not the torsion of an affine connection, but rather the (related) torsion form on the frame bundle, as in Cartan's first structure equation. Since $\W$ is torsion free, it is the unique connection compatible with the frame $\E_\W$. The equation $\F(\W) = \E_\W\wedge \E_\W$ satisfied by the frame is then the constant curvature condition for the metric for $\E_\W$.
We can also describe the relation between the 3D and 6D action functionals.
\begin{proposition}
The value of the Hitchin functional on $\Omega=CS(\W)$ is a constant multiple of the volume of the base, computed in terms of the metric for $E_\W$. The Euler-Lagrange equation obtained by extremising this volume is the torsion-free condition $\nabla_\W \E_\W=0$.
\end{proposition}
This can be paraphrased by saying that the Hitchin action on $\Omega=CS(\W)$ is just a multiple of the pure connection action for 3D gravity. We would like to emphasise that not only does this construction provides a lift of 3D gravity to 6D, but it also explains why the pure connection formulation of 3D gravity is possible, by interpreting it as Hitchin's theory for the Chern-Simons 3-form for the connection $\W$. This description of 3D gravity as Hitchin's theory for $CS(\W)$ in the total space of the bundle $\M\to M$ is one of the main results of the present paper. The 6D interpretation developed here is useful in practice, e.g. for the description of the arising gradient flow, which is a version of the Ricci flow, as will be addressed in another publication.

Given that solutions of 3D gravity can be lifted to 6 dimensions, where they define 3-forms that are critical points of the Hitchin functional, there arises a question if one can interpret 3D gravity as a {\it dimensional reduction} of the 6D Hitchin theory of 3-forms. The other main aim of this paper is to give an (affirmative) answer to this question. Moreover, as we shall see, what arises in the process of dimensional reduction is more than just the theory of 3D gravity. The dimensionally reduced theory is 3D gravity coupled to a (topological) 2-form field.

To carry out the dimensional reduction, we assume that we have some 3-dimensional Lie group $G$ (with $G$ being either ${\rm SU}(2)$ or ${\rm SL}(2,\R)$ depending on the desired 3D metric signature) acting freely on our 6-dimensional manifold $\M$. In this paper we only consider the case $G={\rm SU}(2)$, but the constructions generalise (with appropriate changes) to the case of ${\rm SL}(2,\R)$. Thus, we assume that the 6-dimensional space $\M$ has the structure of the total space of the principal ${\rm SU}(2)$ bundle ${\rm SU}(2)\hookrightarrow \M\to M$ over base $M$. We consider ${\rm SU}(2)$ invariant 3-forms $\Omega$ on $\M$ that are critical points of the Hitchin functional, and interpret the data stored in $\Omega$ in terms of certain fields on the base $M$.  More specifically, we consider 3-forms $\Omega=\Omega_*+d\B$, where $\Omega_*$ is a fixed closed but not exact form on $\M$, and $\B\in \Lambda^2 \M$ is a $G$-invariant 2-form. Fixing $\Omega_*$, we interpret the Hitchin functional as a functional of the 2-form $\B$.

To understand what to expect in terms of fields on the base, it is instructive to count the numbers of components available. The number of components of a general 2-forms in 6D is $15$. However, not all components in the field $\B$ contribute, as any shift by an exact 2-form is irrelevant. In turn, exact 2-forms are only defined modulo exact 1-forms. Thus, the number of relevant components of $\B$ is $15-(6-1)=10$. We show that these components split into $9=3\times 3$, which receive the interpretation of the components of a ${\rm SU}(2)$ connection, plus $1=3-(2-1)$, which is interpreted as a 2-form mod 1-form mod zero-form in 3D. We must however also take into account the gauge symmetries and, after analysing these, one finds that none of these $10$ components describe propagating degrees of freedom.

Thus, to carry the dimensional reduction we parametrise the 3-form $\Omega$ by an ${\rm SU}(2)$ connection and a 2-form on the base. The simplest situation arises when the 2-form is set to zero. This turns out to correspond to the case when $\Omega$ does not have a purely horizontal component. In this simplest case $\Omega$ turns out to be just the Chern-Simons 3-form for the connection $\W$ defined by $J$. The Euler-Lagrange equations of Hitchin's theory then become field equations of 3D gravity in the pure connection formulation, as we explained above.  

The general case when $\Omega$ has also a purely horizontal component is much more intricate. In this case it is difficult to parametrise $\Omega$ by the connection that it defines, and we parametrise $\Omega$ by some ${\rm SU}(2)$ connection and by a 2-form field on the base. One can then compute the connection defined by $\Omega$. The field equations of Hitchin's theory then state that the field strength of the 2-form field on the base is constant, that the connection defined by $\Omega$ is torsion free, and that the metric has constant negative curvature. Together, these are the field equations of 3D gravity coupled to the 2-form field. 

To put the results of this paper in a more general context, we recall that in Kaluza-Klein dimensional reduction a gravitational theory in $D+1$ dimensions gives rise to a $D$ dimensional theory of gravity coupled to extra fields. By assuming invariance under the action of a 1-dimensional group, the components of the $D+1$ metric aquire the interpretation of a metric, a 1-form and a scalar field in $D$ dimensions. Thus, starting with a theory of gravity in $D+1$ dimensions one obtains gravity coupled to electromagnetism and a scalar field in $D$ dimensions.

In this paper we consider an analogous dimensional reduction, but the theory to be reduced is {\it not} a gravitational theory. It is defined in 6 dimensions, and can be thought of as a theory of a single {\it 2-form field}. The action is constructed as a non-linear diffeomorphism invariant functional of the corresponding field strength, which is a 3-form. Prior to dimensional reduction, no metric appears in this theory, not even as a secondary object.

The main result of this paper is that the dimensional reduction to 3 dimensions of this diffeomorphism invariant 6D theory is the theory of 3D gravity, in general coupled to a topological 2-form field. It is a striking result that a gravitational theory in a lower number of dimensions appears as the dimensional reduction of a theory of a very different nature in higher dimensions. In particular, at first sight, apart from being diffeomorphism invariant, the theory being dimensionally reduced seems to have very little to do with gravity.

Let us also remark that another way to see the link between 3D and 6D structures is given via the Bryant-Salamon \cite{BS} construction of $G_2$ holonomy metrics on the 7-dimensional total space of the spinor bundle over a constant curvature 3D manifold. The arrising 3-forms in 7D, defining the $G_2$ structure, can then be obtained as evolving 6D 3-forms in the sense of Hitchin, see Section 6 of \cite{Hitchin:2001rw}. However, the arising 6D almost complex structure in this case is not integrable. So, this 7D interpretation, while not unrelated to the main 6D story, is different. The lift of solutions of 3D gravity to 7D $G_2$ holonomy metrics has also been described in \cite{Dijkgraaf:2004te}. We will describe this 7D interpretation of the 3D metrics below. 

The organisation of this paper is as follows. We start, in Section \ref{sec:hitchin} by a review of Hitchin's 6D theory. We review 3D gravity in Section \ref{sec:3d}. We then review the topological theory of a 2-form field in Section \ref{sec:2-form}. We give the pure connection description of both $\Lambda\not=0$ 3D gravity and gravity coupled to the 2-form field in Section \ref{sec:pure}. The pure connection description, even if not new, is not very-well known, and is essential to understand the constructions of this paper. We describe the lift of the pure connection description of gravity to 6D in section \ref{sec:lift}. It is here that the Chern-Simons form on the total space of the principal ${\rm SU}(2)$ bundle first appears. We then turn to the problem of the dimensional reduction in Section \ref{sec:reduction}. We see how the ansatz of Section \ref{sec:lift} with its Chern-Simons 3-form arises from the dimensional reduction. Here we also describe how certain components of Hitchin's 3-form receive the interpretation of an ${\rm SU}(2)$ connection when the theory is compactified. We show that what arises in the process of dimensional reduction is in general 3D gravity plus the 2-form field. We describe a 7D interpretation of the metric that arises in 6D in Section \ref{sec:7d}. We conclude with a discussion. 

\section{Geometry of 3-forms in six dimensions}
\label{sec:hitchin}

In this section we review the material from \cite{Hitchin:2000jd}, in the amount necessary for our purposes. We shall denote the 6D manifold by $P$ already at this stage, keeping it in mind that later we want this manifold to be the total space of the principal bundle over a 3D base. A 3-form in 6 dimensions {\it does not} define an orientation of the manifold. For this reason, we assume our manifold $\M$ to be orientable, and pick an orientation. It will be necessary for various constructions below, in particular to define the volume functional --- the Hitchin action. 

\subsection{Stable forms over $\C$}

A stable 3-form in six dimensions is one that lies in an open ${\rm GL}(6)$ orbit. For complex 3-forms, there is a single open orbit. It is thus easier to start with the description of the situation over $\C$ and later specialise to the real case.

Over complex numbers, we have the following proposition from \cite{Hitchin:2000jd}. A stable 3-form $\Omega$ is the sum of two decomposable 3-forms 
\be\label{om-canonical}
\Omega=\alpha_1\wedge \alpha_2\wedge \alpha_3 + \beta_1\wedge \beta_2\wedge \beta_3, \qquad \alpha_1\wedge \alpha_2\wedge \alpha_3 \wedge \beta_1\wedge \beta_2\wedge \beta_3 \not=0,
\ee
with 1-forms $\alpha_i, \beta_i, i=1,2,3$ thus forming a basis in $T^*\M^\C$. The decomposable 3-forms $\alpha_1\wedge \alpha_2\wedge \alpha_3,\beta_1\wedge \beta_2\wedge \beta_3$ are unique, and we distinguish between $\alpha_i$ and $\beta_i$ by requiring that $\alpha_1\wedge \alpha_2\wedge \alpha_3 \wedge \beta_1\wedge \beta_2\wedge \beta_3$ is in the fixed orientation class. The component of the stabiliser of (\ref{om-canonical}) in ${\rm GL}(6,\C)$ connected to the identity is the group ${\rm SL}(3,\C)\times{\rm SL}(3,\C)$, with the first of ${\rm SL}(3,\C)$'s permuting the triple $\alpha_i$ and the second acting on $\beta_i$.

Let us now pick an arbitrary volume form $v$ on $\M$, for definiteness in the chosen orientation class. Given a volume form on $\M$, a stable 3-form in six dimensions defines a endomorphism of the (co-) tangent bundle $T^*\M^\C$, that will later be used to define the almost-complex structure. Different volume forms are related by multiplication by a nowhere vanishing function, and the endomorphism will be defined modulo multiplication by such a function. For a 1-form $\alpha$, we define another 1-form $K_\Omega(\alpha)$ via
\be
i_\xi( K_\Omega (\alpha)) := \alpha \wedge i_\xi \Omega \wedge \Omega/ v, \qquad \forall \xi\in T\M.
\ee
It is easy to see that for $\Omega$ given by (\ref{om-canonical}), and using $v= v_\Omega\equiv \alpha_1\wedge \alpha_2\wedge \alpha_3 \wedge \beta_1\wedge \beta_2\wedge \beta_3$, we have
\be\label{k-action}
K_\Omega (\alpha_i) = \alpha_i, \qquad K_\Omega (\beta_i) = -\beta_i.
\ee

The result (\ref{k-action}) of the action of $K_\Omega$ on the basic forms $\alpha_i,\beta_i$ shows that $K^2$ is a multiple of the identity operator
\be
K_\Omega^2 = \lambda(\Omega) \id,
\ee
where
\be\label{lambda-om}
\lambda(\Omega) := \frac{1}{6} {\rm Tr}(K_\Omega^2).
\ee
This is an invariant of $\Omega$ with degree four under ${\rm SL}(6)$ transformations. A 3-form $\Omega$ is stable if and only if $\lambda(\Omega)\not =0$. 

\subsection{Stable forms over $\R$, case $\lambda(\Omega)>0$}

There are exactly two ${\rm GL}(6,\R)$ generic orbits, characterised by the sign of the invariant (\ref{lambda-om}). Note that this sign is invariantly defined, since as we multiply $v\to fv$, where $f$ is a nowhere vanishing function on $\M$ (which can be everywhere negative) the endomorphism transforms as $K_\Omega\to K_\Omega/ f$, but $\lambda(\Omega)\to \lambda(\Omega)/f^2$. 

For the plus sign the 3-form $\Omega$ has the canonical form (\ref{om-canonical}) with $\alpha_i,\beta_i$ real 1-forms. The 3-form $\Omega$ defines the volume form 
\be
v_\Omega := \sqrt{\lambda(\Omega)} v.
\ee
Here $v$ is a top form in the given orientation class. Note that the right-hand-side now does not depend on which volume form $v$ is used to define $K_\Omega$, and is thus a volume form in the fixed orientation class. We can integrate this volume form to get the volume functional
\be
S[\Omega] := \int_\M v_\Omega.
\ee
This is the Hitchin functional for the case $\lambda(\Omega)>0$. 

We also note that there is a constructive way of obtaining the decomposable 3-forms $\alpha_1\wedge \alpha_2\wedge \alpha_3, \beta_1\wedge \beta_2\wedge \beta_3$. Indeed, define the operator
\be
I_\Omega := \frac{1}{\sqrt{\lambda(\Omega)}} K_\Omega.
\ee
This operator is canonically defined in the sense that it does not depend on which volume form $v$ is used in the construction of $K_\Omega$. This operator squares to the identity $I_\Omega^2=\id$, and using it we construct the decomposable forms
\be
2 \alpha_1\wedge \alpha_2\wedge \alpha_3  = \Omega + \hat{\Omega}, \qquad 
2\beta_1\wedge \beta_2\wedge \beta_3 = \Omega - \hat{\Omega},
\ee
where 
\be
\hat{\Omega}:= I_\Omega(\Omega) 
\ee
is the result of the action of $I_\Omega$ on all 3 form indices of $\Omega$. Once the decomposable 3-form factors of $\Omega$ have been obtained, the volume form $v_\Omega$ defined by $\Omega$ is given by
\be
v_\Omega = \alpha_1\wedge \alpha_2\wedge \alpha_3 \wedge \beta_1\wedge \beta_2\wedge \beta_3 =\frac{1}{2} \hat{\Omega}\wedge \Omega.
\ee
Finally, we note that for this positive sign of $\lambda(\Omega)$, the stabiliser in ${\rm GL}(6,\R)$ of $\Omega$ is ${\rm SL}(3,\R)\times{\rm SL}(3,\R)$, each of which acts on its own triple $\alpha_i,\beta_i$. 

\subsection{Stable forms over $\R$, case $\lambda(\Omega)<0$}

The negative case sign $\lambda(\Omega)<0$ is more interesting. In this case, the canonical form for $\Omega$ is
\be\label{om-complex}
\Omega = \alpha_1\wedge \alpha_2 \wedge \alpha_3 + \bar{\alpha}_1\wedge \bar{\alpha}_2\wedge \bar{\alpha}_3, \qquad \alpha_1\wedge \alpha_2 \wedge \alpha_3 \wedge \bar{\alpha}_1\wedge \bar{\alpha}_2\wedge \bar{\alpha}_3 \not=0,
\ee
where $\alpha_i$ are now complex-valued 1-forms, and $\bar{\alpha}_i$ are the complex conjugate forms. The distinction between $\alpha_i$ and the complex conjugate forms is obtained by requiring that $\im \alpha_1\wedge \alpha_2 \wedge \alpha_3 \wedge \bar{\alpha}_1\wedge \bar{\alpha}_2\wedge \bar{\alpha}_3$ is in the fixed orientation class. 

There are several alternative ways of writing the canonical form of $\Omega$. We can write
\be\label{omega-real-part}
\Omega = 2\,{\rm Re}(\alpha_1\wedge \alpha_2 \wedge \alpha_3).
\ee
The stabiliser in this case is ${\rm SL}(3,\C)$. Yet another way of writing $\Omega$ in this case arises if we set $\alpha_1 = w_1+\im e_1$, etc., where $w_i,e_i$ are real 1-forms. We get
\be
\Omega/2 = w_1\wedge w_2\wedge w_3 - w_1\wedge e_2\wedge e_3 - w_2\wedge e_3\wedge e_1-w_3\wedge e_1\wedge e_2.
\ee

In analogy to what we have done for the positive sign, we define the operator
\be\label{ACS}
J_\Omega := \frac{1}{\sqrt{-\lambda(\Omega)}} K_\Omega,
\ee
which is invariantly defined, i.e. is independent of the choice of $v$ in the definition of $K_\Omega$. The operator defined has the property
\be
J_\Omega^2 = -\id,
\ee
and so defines an almost-complex structure on $\M$. We have the following action on the basic forms
\be\label{J-alpha}
J_\Omega (\alpha_i) = \frac{1}{\im} \alpha_i,
\ee
so that $\alpha_i$ are $(0,1)$-forms.

We also have an invariantly defined volume form
\be
v_\Omega := \sqrt{-\lambda(\Omega)} v,
\ee
independent of the choice of $v$ (in the fixed orientation class). For $\Omega$ written in its canonical form (\ref{om-complex}) we get
\be
v_\Omega = \im\, \alpha_1\wedge \alpha_2 \wedge \alpha_3 \wedge \bar{\alpha}_1\wedge \bar{\alpha}_2\wedge \bar{\alpha}_3.
\ee

We can get the decomposable factors $\alpha_1\wedge \alpha_2 \wedge \alpha_3$ and its complex conjugate by acting with $J_\Omega$ on $\Omega$ and adding (or subtracting) the result to $\Omega$. Thus, we define
\be
\hat{\Omega} := - J_\Omega ( \Omega),
\ee
where the sign is chosen for convenience. We then have
\be\label{omega-c}
\Omega^c\equiv \alpha_1\wedge \alpha_2 \wedge \alpha_3 = \frac{1}{2} (\Omega + \im \hat{\Omega}),
\ee
and
\be\label{v-omega*}
v_\Omega = \frac{1}{2} \Omega\wedge \hat{\Omega}.
\ee

\subsection{The volume functional and its critical points}

We now concentrate on the more interesting negative sign case $\lambda(\Omega)<0$. The 3-form $\Omega$, together with the chosen orientation of $\M$, define the volume functional
\be\label{action-hitchin}
S[\Omega]=\int_\M v_\Omega.
\ee
In what follows, we will need its first variation which is easily obtained from (\ref{v-omega*}), see \cite{Hitchin:2000jd}. We have
\be
\delta S[\Omega] = \int_\M \delta \Omega \wedge \hat{\Omega}.
\ee

We can then consider variations of the 3-form $\Omega$ staying within a fixed cohomology class
\be
\Omega = \Omega_* + dB, \qquad d\Omega_*=0, \quad B\in \Lambda^2\M.
\ee 
In particular, we have $d\Omega=0$ and consider variations of the form $\delta\Omega = d\delta B$. The resulting Euler-Lagrange equations are then given by
\be
d\hat{\Omega} = 0.
\ee
Thus the $(3,0)$-form $\Omega^c$ is closed, which in turn implies that the almost-complex structure $J_\Omega$ is integrable and so $\M$ is a complex manifold, see \cite{Hitchin:2000jd}. 

\subsection{Circle action}

The space of 3-forms in 6D has a natural symplectic structure. Its value on two tangent vectors $\delta_1\Omega,\delta_2\Omega$ is
\be
\omega(\delta_1\Omega,\delta_2\Omega)=\int_\M \delta_1\Omega\wedge \delta_2\Omega.
\ee
One can then view the volume functional $S[\Omega]$ defined in (\ref{action-hitchin}) as a Hamiltonian function. This Hamiltonian generates on the space of 3-forms a circle action, see section 2.9 of \cite{Hitchin:2000jd}. A finite version of the arising transformation on the space of 3-forms is easiest described in terms of the complex 3-form $\Omega^c$ of which $\Omega$ is (twice) the real part. It turns out that the circle action on $\Omega^c$ is given simply by multiplication $\Omega^c\to e^{\im\theta}\Omega^c$. In real terms we have
\be\label{circle}
\Omega \to \cos(\theta)\Omega - \sin(\theta)\hat{\Omega},
\ee
where $\theta$ is the (finite) transformation parameter. We remark that the complex structure defined by $\Omega$ stays unchanged when the $\Omega$ is transformed this way. It is also obvious that the Hitchin action (\ref{action-hitchin}) is invariant under (\ref{circle}). 

We will use this circle action as a trick to put a given 3-form into the canonical form in Section \ref{sec:reduction}. 

\section{3D Gravity with cosmological constant}
\label{sec:3d}

In this section, we review some basic facts about 3D gravity. Our notations are standard for the gravity literature. We also describe here a somewhat less known 3D gravity deformed by a "topological term".

\subsection{Einstein-Cartan frame formalism in 3D}

Let $e^i, i=1,2,3$ be a frame field so that the 3D metric is 
\be
ds^2 =e^i \otimes e^j \eta_{ij},
\ee
where $\eta_{ij}$ is either $\eta_{ij}={\rm diag}(1,1,1)$ or $\eta_{ij}={\rm diag}(-1,1,1)$ depending on the desired signature. There are subtle differences between the two signature cases. For definiteness, let us consider the all plus signature case, which is also what we need for the most interesting application to hyperbolic manifolds. 

For the Riemannian signature we raise and lower indices with the metric $\delta_{ij}$, and the ${\rm SO}(3)$ spin connection is the set of 1-forms $w^{ij}=w^{[ij]}$. The anti-symmetry is the statement that the connection is $\delta_{ij}$ metric compatible. Let $f^{ij}$ be the curvature
\be
f^{ij} = dw^{ij} + w^{ik} \wedge w_k{}^j.
\ee
We then write the following action
\be\label{s-wt}
S[e,w]= - \frac{1}{4} \int_M \left( e^i \wedge f^{jk}  - \frac{\Lambda}{3} \, e^i\wedge e^j\wedge e^k \right) \epsilon_{ijk} .
\ee
The orientation implied here is that of the 3-form $e^i\wedge e^j\wedge e^k \epsilon_{ijk}$. The minus sign in front of the action is the usual choice for the all plus signature. We work in units in which the 3D Newton's constant satisfies $4\pi G=1$. Varying this action with respect to $w$ we get the torsion-free condition
\be\label{torsion-free}
\nabla_w e^i \equiv de^i + w^i{}_j \wedge e^j=0.
\ee
It says that the connection $w$ is the unique $e$-compatible connection. Substituting this connection into (\ref{s-wt}) we find
\be
S[e,w(e)] = - \frac{1}{4}\int_M (R-2\Lambda) v_g,
\ee
where $R$ is the Ricci scalar of the metric, and the integration is carried out with respect to the metric volume element $v_g$. 

The connection matrix $w^{ij}$ being anti-symmetric, we can write 
\be
w^{ij} = \epsilon^{ikj} w^k,
\ee
which defines the new connection 1-forms $w^i$. We then have for the curvature 
\be
f^{ij} = \epsilon^{ikj} f^k, \qquad f^i = dw^i + \frac{1}{2} \epsilon^{ijk} w^j \wedge w^k.
\ee

\subsection{Matrix notations}

It is very convenient to get rid of the internal $i,j,\ldots$ indices at the expense of making all objects $2\times 2$ matrix valued. To this end, we use the isomorphism of the Lie algebras ${\mathfrak so}(3)={\mathfrak su}(2)$. The Lie algebra generators are 
\be\label{tau}
\tau_i = -\frac{\im}{2} \sigma_i,
\ee
where $\sigma_i$ are the usual Pauli matrices. We have
\be
{\rm Tr}(\tau_i \tau_j) = - \frac{1}{2} \delta_{ij}, \qquad
[\tau_i, \tau_j]  = \epsilon_{ij}{}^k \tau_k.
\ee
The index of $\epsilon$ here is raised with the $\delta^{ij}$ metric. 

We then form a matrix-valued connection
\be
\w := w^i \tau_i.
\ee
In what follows we will always denote a matrix-valued object by a bold-face letter. The matrix valued curvature $\f:= f^i \tau_i$ is computed as
\be
\f = d\w + \w\wedge \w.
\ee
We also form anti-hermitian frame field 1-forms
\be
\e: = e^i \tau_i,
\ee
in terms of which the metric is
\be
ds^2 = - 2\, {\rm Tr} (\e\otimes \e).
\ee
In terms of the matrix-valued fields the torsion-free condition (\ref{torsion-free}) takes the form
\be\label{nabla}
\nabla\e \equiv d\e + \w\wedge \e + \e\wedge \w = 0.
\ee
The field equation obtained by varying the action (\ref{s-wt}) with respect to $\e$ takes the form
\be
\f=-\Lambda \e\wedge \e.
\ee
The action itself takes the form
\be\label{s-matrix}
S[\e,\w] = - \int_M {\rm Tr}\left( \e\wedge \f + \frac{\Lambda}{3} \e\wedge\e\wedge \e\right).
\ee
In what follows, we will mainly consider the $\Lambda<0$ case. We set for simplicity
\be
\Lambda = -1.
\ee
A different value of $|\Lambda|$ can always be reinstalled by rescaling the frame field.  

\subsection{Chern-Simons formulation}

The two sets of equations $\nabla \e=0, \f=\e\wedge\e$ can be combined as the real and imaginary parts of a single complex-valued equation by introducing the complex tracefree $2\times 2$ matrix-valued field
\be\label{CS-connection}
\A:= \w + \im \e.
\ee
The field equations of 3D gravity then combine into the statement that the curvature of the ${\rm SL}(2,\C)$ connection $\A$ is zero
\be
0=\f(\A) \equiv d\A + \A\wedge \A.
\ee
These are the field equations following from the Chern-Simons Lagrangian. Alternatively, we can write the Einstein-Cartan Lagrangian (\ref{s-matrix}) (with $\Lambda=-1$), modulo a surface term, as
\be\label{grav-cs}
S[\e,\w]= - \frac{1}{2}  {\rm Im} \int_M  CS[\A] ,
\ee 
where
\be
CS[\A]:= {\rm Tr}\left( \A\wedge d\A + \frac{2}{3} \A\wedge \A\wedge \A\right)
\ee
is the Chern-Simons 3-form for $\A$. 

\subsection{Topological term}

It is possible to add to (\ref{s-matrix}) also the real part of the Chern-Simons functional of $\A$ with an arbitrary coefficient, see \cite{Witten:1988hc}, section 2.3.  When written in terms of $\e,\w$ this reads
\be\label{s-top}
 {\rm Re} \int_M  CS[\A]  = \int_M \left( CS[\w] - {\rm Tr}(\e \nabla \e)\right).
\ee
It is not hard to check that this term does not affect the field equations, in the sense that a linear combination of the two resulting field equations still says that the connection is metric compatible. 

\subsection{Torsion}

In order to better understand what happens in the 6D story we need to consider a certain other "deformation" of the usual 3D gravity action. This deformation does affect field equations and renders the connection {\it not} metric compatible. Thus, it introduces torsion. Let us see how this works.

We shall add to the (\ref{s-matrix}) just the second term from (\ref{s-top}) with an arbitrary coefficient. Thus, we consider the following deformation
\be\label{s-kappa}
S[\e,\w]= -\int_M {\rm Tr}\left( \e \f - \frac{1}{3}\e^3 - \kappa \,\e\nabla \e\right),
\ee
where $\kappa\in \R$ is a new coupling constant. This is still an action containing at most first derivatives of the basic fields, and so is an acceptable construct. Note that the term we have added is gauge invariant, unlike (\ref{s-top}) that is not invariant under the so-called large gauge transformations. Let us see how the presence of this term affects the field equations. Varying this action with respect to the connection gives
\be\label{kappa-w}
\nabla \e - 2\kappa\,  \e\wedge \e=0.
\ee
Varying with respect to the frame field gives
\be\label{kappa-e}
\f - \e\wedge \e - 2\kappa \nabla \e=0.
\ee

The equation (\ref{kappa-w}) says that the connection $\w$ is not metric compatible. Let us represent $\w$ as the sum of the metric compatible connection and the torsion
\be\label{w-t}
\w=\w_\e +\bt,
\ee
where by definition
\be
d\e+\w_\e \e+\e \w_\e \equiv \nabla_{\w_\e} \e =0.
\ee
Using this representation of $\w$ the equation (\ref{kappa-w}) becomes
\be
\bt \e +\e\bt = 2\kappa\, \e\e.
\ee
It is clear that a solution of this equation is
\be\label{torsion}
\bt = \kappa \e.
\ee
It is also easy to check that this is the unique solution. 

Let us now interpret the second equation (\ref{kappa-e}). Using (\ref{kappa-w}) we can rewrite (\ref{kappa-e}) as
\be
\f(\w) = (1+4\kappa^2) \e\e.
\ee
Let us now compute the curvature $\f$ using the representation (\ref{w-t}). We have
\be
\f(\w) = \f(\w_\e) + \kappa \nabla_{\w_\e} \e + \kappa^2 \e\e.
\ee
Using the fact that the second term here is zero, and combining the the above two equations we get
\be
\f(\w_\e) = (1+3\kappa^2) \e\e.
\ee
Thus, the role of the deformation term in (\ref{s-kappa}) is to make the connection that appears in the action to have torsion. Once the torsion is solved for, one finds the metric to still have constant negative curvature, with the effective cosmological constant
\be\label{lambda-kappa}
\Lambda_{eff} = -(1+3\kappa^2).
\ee

\subsection{Field redefinition}

It is possible to derive the results of the previous subsection in a simpler way by a field redefinition. Indeed, let us introduce a new connection $\tilde{\w}$ so that
\be
\w = \tilde{\w} +\kappa\e.
\ee
We can the write the action (\ref{s-kappa}) in terms of $\tilde{\w}$. The terms $\e\tilde{\nabla}\e$ then cancel and we get
\be
S[\e,\tilde{\w}] = -\int_M {\rm Tr}\left( \e\tilde{\f} - \frac{1+3\kappa^2}{3} \e^3\right).
\ee
Here $\tilde{\f}$ is the curvature of $\tilde{\w}$. This is the usual first order action of 3D gravity with the effective cosmological constant (\ref{lambda-kappa}). 

Thus, the torsion term added in (\ref{s-kappa}) can be eliminated by a simple redefinition of the connection. The redefined connection is then torsion free as the consequence of its field equation. Below we shall see that a very similar mechanism is at play in the dimensional reduction of 6D Hitchin theory to 3D. 

\subsection{Quantum theory}

Even though this paper is about classical theory, it is appropriate to give some comments on the quantum case. The theory of 3D gravity being topological, one expects to be able to construct the corresponding quantum theory. The easiest case is that of $\Lambda=0$. In this case, for Riemannian signature, the theory becomes what is known as ${\rm SU}(2)$ $BF$ theory. It is one-loop exact, and the partition function can be explicitly computed. It reduces to the Ray-Singer torsion for the operator $\nabla$, see e.g. \cite{Birmingham:1991ty}. 

The case of non-zero $\Lambda$ is much harder, as the theory is no longer one-loop exact. In spite of this, the quantum version of the Riemannian signature $\Lambda>0$ gravity is known. It is based on the quantum group ${\rm SU}_q(2)$ (at root of unity), see \cite{Reshetikhin:1991tc}. The partition function on a given 3-manifold $M$ is constructed by choosing a simplicial decomposition of $M$, and then decorating the arising simplicial complex with certain combinatorial data. The arising state sum is independent of the chosen simplicial decomposition and is a topological invariant of $M$. At least in part of the literature, this construction is referred to as the Turaev-Viro model. Another way of seeing why the $\Lambda>0$ case is understood is by noticing that in this case the Lagrangian can be represented as the difference of two Chern-Simons Lagrangians for $\w\pm\e$. The quantum Chern-Simons theory for the gauge group ${\rm SU}(2)$ is understood, and in a precise sense the $\Lambda>0$ 3D gravity partition function is the product of two CS partition functions, see e.g. \cite{Roberts} for a nice proof. 

As far as we are aware, there is no complete construction of the much more difficult $\Lambda<0$ quantum theory, even though there is some recent progress in this direction, see e.g. \cite{Blau:2016vfc} and references therein. 

\section{2-form field}
\label{sec:2-form}

In this section we review the theory of a 2-form field in 3D. We need this material because, as we shall see below, the dimensional reduction of the 6D theory produces 3D gravity coupled to such a field. 

Consider a theory of $(d-1)$-form in $d$ dimensions, with the Lagrangian being the corresponding field strength squared. As is well-known, this theory is topological because the field equation for the $(d-1)$-form implies that its field strength is a constant. This constant shifts the value of any otherwise present cosmological constant, see e.g. \cite{Duff:1980qv} for the description of this theory in 4D. Let us give a brief description of the 3D situation. We first give a local analysis, and then analyse the global subtleties that may be present. 

\subsection{The theory}

One can add to 3D gravity a field that does not describe any propagating degrees of freedom. This is achieved as follows. Consider a 2-form field $b\in \Lambda^2M$ with the following action
\be\label{s-B}
S[b]= \frac{1}{2}\int_M v_g (db)^2,
\ee
where $(db)^2$ is computed using the metric, and $v_g$ is the metric volume element. The field equations one gets by varying this Lagrangian with respect to $b$ are
\be\label{B-feqs}
d*(db)=0
\ee
This says that $*(db)$, the Hodge dual of the 3-form $db$, is a constant. There is a subtlety that on a compact manifold $M$ this constant must be zero, see below. We will come back to this global issue below. For now we continue with a local analysis. 

\subsection{The partition function}

There are many different ways to see that this theory does not have propagating degrees of freedom. One way is to compute the path integral and see that all the arising determinants cancel. To this end, we introduce the parametrisation
\be\label{B-param}
b= d\theta + *(d\phi),
\ee
where $\theta\in \Lambda^1M$ and $\phi$ is a function. Here we are only interested in demonstrating that there are no propagating degrees of freedom, and so we have ignored (a finite number of) harmonic terms that may be present. The quantum mechanics of these modes needs to be analysed separately. In this parametrisation the Lagrangian in (\ref{s-B}) gives $(1/2)\phi \Delta_0^2 \phi$, and so the path integral gives $1/{\rm det}(\Delta_0)$, where $\Delta_0= *d*d$. But there is also the Jacobian related to the chosen parametrisation of the field $b$. A part of this Jacobian that has to do with the second term in (\ref{B-param}) is ${\rm det}(*d)=\sqrt{{\rm det}(*d*d)}=\sqrt{{\rm det}\Delta_0}$. But there is also the part that arises from the first term in (\ref{B-param}). This parametrisation by 1-forms $\theta$ is modulo exact one-forms. Because of this to compute the determinant one needs ghosts. This gives for this part of the Jacobian $\sqrt{{\rm det}\Delta_1}/{\rm det}\Delta_0$, where $\Delta_1$ is the Laplacian one 1-forms. Collecting all the factors we see that the path integral over $b$ for theory (\ref{s-B}) gives
\be\label{ray-singer}
\int {\cal D}b \, e^{-S[b]} = \left(\frac{{\rm det}\Delta_1}{({\rm det}\Delta_0)^3}\right)^{1/2}.
\ee
One recognises in this quantity the Ray-Singer torsion $T(3)$, see e.g. \cite{Birmingham:1991ty}, section 6.2.1. It is a topological invariant of $M$, i.e. does not depend on the metric used to write (\ref{s-B}). This expression for the partition function also shows how there are no degrees of freedom in this theory. Indeed, for purposes of counting the DOF, the dimension of the space of 1-forms is 3, and thus each eigenvalue of $\Lambda_1$ has multiplicity 3, which is cancelled by the third power of eigenvalue of $\Lambda_0$ in the denominator. 

\subsection{Hamiltonian analysis}

There is a simpler argument to show the absence of the propagating DOF which proceeds by the Hamiltonian analysis of (\ref{s-B}). For simplicity, let us do this analysis in flat space. Performing the space plus time decomposition, the Lagrangian is
\be
\frac{1}{6} (\dot{b}_{ab} + 2\partial_{[a} b_{b]0}) \dot{b}^{ab},
\ee
where $a,b$ are the spatial indices. The only field here with time derivatives is $b_{ab}$, the spatial projection of the 2-form $b$. The corresponding conjugate momentum is
\be
p^{ab} = \frac{\partial L}{\partial \dot{b}_{ab}} = \frac{1}{3} (\dot{b}_{ab} + 2\partial_{[a} b_{b]0}).
\ee
The Hamiltonian is $H = p^{ab} \dot{b}_{ab} - L$ and is equal to
\be
H = \frac{1}{2} (p^{ab})^2 - \frac{1}{3} p^{ab} \partial_a b_{b0}.
\ee
We see that the field $b_{a0}$ is a Lagrange multiplier imposing the constraint $\partial_a p^{ab}=0$. But since on the 2-dimensional spatial slice $p^{ab} =\epsilon^{ab} p$ for some scalar field $p$, we see that the constraint says $\partial_a p=0$, and thus the momentum is a constant. The constraint generates gauge transformations $\delta b_{ab} = \partial_{[a} \theta_{b]}$, where $\theta_a$ is the generator of the transformation. It is clear that this transformation can be used to set the field $b_{ab}=\epsilon_{ab} b$ to a constant. Thus, the reduced phase space is zero dimensional and there are no propagating DOF. 

\subsection{Action in terms of frame field}

For what follows, it will also be useful to write the action (\ref{s-B}) using the frame fields. In terms of the frame the volume element is $v_g=-(2/3){\rm Tr}(\e^3)$. We can then introduce the notation
\be
\chi: = *(db) = db/v_g.
\ee
 The action (\ref{s-B}) is then written as
\be\label{SB}
S[b] = -\frac{1}{3} \int_M \chi^2 {\rm Tr}(\e\wedge \e\wedge \e).
\ee
Its variation with respect to $b$ gives the equation that says that $\chi$ is a constant. 

\subsection{3D gravity coupled to the 2-form field}

Let us now add together the terms (\ref{s-matrix}) and (\ref{SB}) to form the action of 3D gravity theory coupled to the 2-form field
\be
S[\e,\w,b]= -\int_M {\rm Tr}\left(\e\f-\frac{1}{3}(1-\chi^2)\e^3 \right).
\ee
Let us analyse what the field equations imply. The variation with respect to $\w$ is unchanged from what we have considered before, and is the statement that $\w$ is metric compatible. To obtain the variation with respect to $\e$ we need the variation of $\chi$ as we vary $\e$. We have
\be
\delta \chi = - \frac{db}{v_g^2} \delta v_g = - \chi \frac{\delta v_g}{v_g}.
\ee
We also have $\delta v_g = -2 {\rm Tr}(\delta \e \wedge \e\wedge \e)$. Taking this into account we get the following equations for $\e$
\be\label{f-e-chi}
\f-(1+\chi^2) \e\e =0.
\ee
The pre factor that appears on the right-hand-side plays the role of the effective (negative) cosmological constant in this model. We see that its value depends on $\chi$. 

\subsection{Subtleties in the compact case}

So far we have ignored the global issues. Let us now assume that the manifold $M$ is compact. In this case the third cohomology group of $M$ is non-trivial $H^3(M)=\R$. An element of $H^3(M)$ is characterised by its integral over $M$. Because of this, as a representative in $H^3(M)$ one can take a suitable constant multiple of the volume form $v_g$. 

The reason why we need to worry about $H^3(M)$ is that, as described above, the field equation of the two-form theory says  
\be
db = \chi v_g
\ee
with $\chi=const$. But integrating both sides of this equation over $M$ we conclude that the only possible value of this constant is zero.\footnote{We thank Joel Fine for pointing out this subtlety to us.}

To rectify this, we change the action of the two-form theory by introducing a coupling to third cohomology. Thus, let $\omega, d\omega=0$ be a representative of the third cohomology of $M$. Consider the following action
\be\label{s-bomega}
S[b,\omega] = \frac{1}{2} \int_M v_g (db+\omega)^2 .
\ee
This action only depends on the class of $\omega$ as changes of the representative correspond to adding an exact form to $\omega$, which can be absorbed in a redefinition of $b$. Now, as we discussed above, a representative can always be chosen to be a constant multiple of the volume form
\be\label{omega-v}
\omega = c \, v_g.
\ee
We can also write the action as
\be\label{s-bc}
S[b,c] = \frac{1}{2} \int_M v_g \chi^2,
\ee
where 
\be\label{chi-omega}
\chi := (db+\omega)/v_g.
\ee

Now, varying (\ref{s-bomega}) with respect to $b$ we get the following field equation
\be
d*(db+\omega)=0.
\ee
This just says that $\chi$ as introduced in (\ref{chi-omega}) is constant $\chi=const$. Alternatively, using (\ref{omega-v}) we get
\be
db +c\, v_g = \chi\, v_g.
\ee
Now integrating both sides of this equation over $M$ gives
\be
\chi =c.
\ee

Regarding the field equations in the gravitational sector, all of the steps of the previous subsection that lead to (\ref{f-e-chi}) remain unchanged. Thus, we learn that on a compact manifold the shift of the cosmological constant provided by the two-form field is determined by the cohomology class of the 3-form $\omega$ that the 2-form is coupled to in (\ref{s-bomega}). 

\subsection{Remark about the quantum theory}

We have seen that, at least in the classical theory, the 2-form field gives rise to a shift of the cosmological constant. One can therefore decide to assign all of the value of the cosmological constant to the 2-form field, thus setting $\Lambda=0$ from the start. The corresponding gravity action is then that of ${\rm SU}(2)$ $BF$ theory coupled to the 2-form field. One could then attempt to compute the partition function by first integrating out the 2-form field. When the action is (\ref{s-B}), as we have explained after (\ref{ray-singer}), the result is independent of the metric and can be taken out of the integral over $\e,\w$. Then the path integral over metrics would be that of the $BF$ theory. In particular, the result would be $\Lambda$-independent. 

However, this is no longer so when there is a coupling to the third cohomology as in (\ref{s-bomega}). This explains the following puzzle. In contrast to what happens in the $\Lambda=0$ quantum gravity, in the non-zero $\Lambda$ case $\Lambda$ is a parameter in the resulting quantum theory, and all topological invariants one gets depend on it. If we could assign all value of $\Lambda$ to the 2-form field, and carry out the path integral over $b$ to obtain a result that is independent of the metric, it would appear that the full quantum theory is $\Lambda$-independent, which is wrong. The coupling to the 3-form $\omega$ in (\ref{s-bomega}) makes the path integral over the scalar field dependent on the metric as well as the value of the constant $c$ that plays the role of the effective cosmological constant. This makes the full partition function of the model $\Lambda$ dependent, as it should be. 

\section{The pure connection formulation}
\label{sec:pure}

In this section we review the pure connection description of 3D gravity. As far as we are aware, the pure connection formulation of 3D gravity was first worked out in \cite{Peldan:1991mh}, starting from the Hamiltonian point of view. A simpler description, directly at the level of the Lagrangian, appears in Section 3.4 of \cite{Peldan:1993hi}. We will only give the Lagrangian description. 

We consider the case of negative cosmological constant $\Lambda=-1$, and first consider pure gravity, but then generalise to the case when the 2-form field $b$ is present. The idea is start with the first-order Einstein-Cartan action (\ref{s-matrix}), and solve the equation $\f=\e\wedge \e$ for $\e$ as a function of $\f$, substituting the result back into the action.  

To describe the solution, we introduce the notion of definiteness and sign of a connection. These notions resemble those of stability and the sign of a 3-form in the 6D story considered in section \ref{sec:hitchin}. In this section this remains just a resemblance. We will see in Section \ref{sec:lift} that there is much more to this similarity, and the 3D notions introduced here {\it are} the notions of the 6D story. In particular, the notion of definiteness here becomes the notion of stability of the corresponding Chern-Simons form in the total space of the bundle. Similarly, the notion of negative definite connection coincides with the notion of 3-forms of negative type for the corresponding Chern-Simons form. The exposition in this section is to a large extent original.

\subsection{Definite connections}

Let $\w$ be a set of $2\times 2$ anti-hermitian matrix-valued one-forms on $M$, i.e. an ${\rm SU}(2)$ connection. Let $\f=d\w+\w\wedge \w$ be the curvature 2-forms. Let us pick an orientation on $M$. Then, for any volume form $v$ in the fixed orientation class, we define a map from the set of 1-forms to the Lie algebra 
\be\label{phi-def}
\phi_\f: T^*M \to {\mathfrak su}(2), \qquad \phi_\f(\alpha) := \alpha\wedge \f/v.
\ee
This is a map from the 3-dimensional space $T^*M$ to the 3-dimensional Lie algebra ${\mathfrak su}(2)$. We call a connection $\w$ {\it definite} or non-degenerate if this map is an isomorphism. 

For a definite connection, we can construct a certain invariant from its curvature. Thus, consider
\be\label{lambda-f}
\lambda(\f):= \frac{4}{3}{\rm Tr}\left( \phi_\f\otimes \phi_\f (\f) \right).
\ee
The notation here is that $\phi_\f$ acts on both form indices of $\f$, and we have a product of 3 Lie algebra elements under the trace. Note that the sign of $\lambda(\f)$ is invariantly defined. Indeed, if we change the orientation by sending $v\to -v$, the sign of (\ref{lambda-f}) does not change. A connection $\w$ is definite if and only if its curvature satisfies $\lambda(\f)\not=0$.

In this paper we are mainly interested in the case when $\lambda(\f)<0$. This corresponds to the negative cosmological constant case $\Lambda<0$. We will refer to such connections as {\it negative definite}. In this case, the connection defines a frame field $\e_\f$ such that 
\be\label{f-theta-eqn}
\f=\e_\f\wedge\e_\f.
\ee

To see that the above form of $\f$ indeed corresponds to $\lambda(\f)<0$ we can substitute (\ref{f-theta-eqn}) into (\ref{lambda-f}) and compute the sign. This computation is easy if we first compute the action of $\phi_\f$ with $\f$ given by (\ref{f-theta-eqn}) on the frame fields. Thus, let us write $\alpha=\alpha^i e^i$ for some choice of the coefficients $\alpha^i$. Using $\e=e^i\tau^i$ and the algebra of $\tau^i$ we can write the curvature as $\f=(1/2)\epsilon^{ijk} e_\f^i\wedge e_\f^j \tau^k$. To compute $\phi_\f$ let us divide in (\ref{phi-def}) by the volume form for the frame field $e^i_\f$, which is given by $v_\e= (1/6)\epsilon^{ijk} e^i\wedge e^j\wedge e^k$. We get the following result for the map $\phi_\f$
\be
\phi_\f(\alpha) = \alpha^i \tau^i.
\ee
In other words, if the curvature is as in (\ref{f-theta-eqn}) and we divide by the frame volume form in (\ref{phi-def}), then the map $\phi_\f$ takes the frame $e^i_\f$ into the generator $\tau^i$. It is then easy to see that $\lambda(\f)$ for $\f$ as in (\ref{f-theta-eqn}) and with the volume form for $\e_\f$ used in the definition of $\phi_\f$ equals to minus one $\lambda(\f)=-1$. 

To prove that for $\lambda(\f)<0$ the curvature can be written in the form (\ref{f-theta-eqn}) we will describe the corresponding frame field explicitly in the next subsection. 

\subsection{The pure connection formulation}

Consider a negative definite connection $\w$. The volume form
\be
v_\f: = \sqrt{-\lambda(\f)} \, v,
\ee
which is in the fixed orientation class, does not depend on the choice of the volume form $v$ used in its construction. It is thus invariantly defined by the negative definite connection $\w$ and the fixed orientation of $M$. The pure connection formulation gravity action is just the total volume
\be\label{pure-conn-action}
S[\w] =  \int v_\f.
\ee

We can now describe $\e_\f$ that solves (\ref{f-theta-eqn}). It is obtained via the following construction
\be\label{theta-f}
i_\xi \e_\f = (\f\wedge i_\xi \f - i_\xi \f \wedge \f)/2v_\f, \qquad \forall \xi\in TM.
\ee
The matrix on the right-hand-side is anti-hermitian, as the commutator of two anti-hermitian matrices. The frame $\e_\f$ defines the metric $ds^2_\f:= - 2\, {\rm Tr}(\e_\f\otimes \e_\f)$, which is of Riemannian signature. The frame $\e_\f$ has the property that 
\be\label{volume-f}
v_\f = -\frac{2}{3} {\rm Tr}(\e_\f\wedge \e_\f\wedge \e_\f).
\ee
Note that the action (\ref{pure-conn-action}) is just the value of the first-order action (\ref{s-matrix}) on the solution (\ref{theta-f}) of (\ref{f-theta-eqn}). 

\subsection{The first variation and Euler-Lagrange equations}

The expression (\ref{volume-f}) makes it clear that the first variation of the pure connection action is given by
\be\label{var-vf}
\delta S[\w]=- \int 2\,{\rm Tr}(\delta \e_\f \wedge \e_\f\wedge \e_\f) =-
 \int {\rm Tr}(\delta (\e_\f \wedge \e_\f) \wedge \e_\f) =-
 \int {\rm Tr}(\delta \f \wedge \e_\f).
\ee
This shows that the critical points of the pure connection action are connections satisfying the following second-order PDE
\be\label{pure-conn-feqs}
\nabla \e_\f = 0,
\ee
with $\nabla$ given by (\ref{nabla}). This equation says that the connection $\w$ is the unique torsion-free metric connection compatible with the frame $\e_\f$. The equation (\ref{f-theta-eqn}) that defines $\e_\f$ then becomes the statement that the metric constructed from $\e_\f$ is of constant negative curvature. This shows that (\ref{pure-conn-action}) is indeed the pure connection formulation of 3D gravity (with negative $\Lambda$).

\subsection{Hamiltonian formulation}

In the pure connection formulation of 3D gravity, the only variable in the theory is an ${\rm SU}(2)$ connection. It is always possible to consider 3D manifolds of the form $M=\Sigma\times \R$, where $\Sigma$ is some 2D surface. The theory then admits the Hamiltonian formulation. The phase space is parametrised by the restriction of the connection $\w$ to $\Sigma$. The temporal component of the connection is a Lagrange multiplier enforcing the constraint, which is simply the spatial projection of the equation $\nabla \e_f=0$. The other two components of this equation are the evolution equations for the spatial connection. 

More explicitly, the $2+1$ decomposition of the action (\ref{pure-conn-action}) reads
\be\label{action-ham-form}
S[\w] =-2\int dt \int_\Sigma d^2x \, \tilde{\epsilon}^{ab} \,{\rm Tr}\left( (\dot{\w}_a - \nabla_a \w_0) \e_b\right).
\ee
Here $\tilde{\epsilon^{ab}}$ is the spatial completely anti-symmetric tensor density (which is reflected by the tilde above the $\epsilon$ symbol). Indices $a,b$ are those on $\Sigma$, i.e. the spatial ones, and $\nabla_a$ is the covariant derivative with respect to the spatial connection $\w_a$. The object $\e_a$ here should be interpreted as a function of $\dot{\w}_a,\w_a, \nabla_a\w_0$. It is to be obtained from the following equations
\be\label{e-a}
\e_0\e_a -\e_a \e_0 = \dot{\w_a} - \nabla_a \w_0, \qquad \f_{ab} = \e_a\e_b - \e_b\e_a.
\ee
The action is thus a complicated non-linear functional of $\dot{\w}_a, \nabla_a \w_0$ and the spatial curvature $\f_{ab}$. 

The canonically conjugate momentum can be read either from (\ref{action-ham-form}) or from (\ref{var-vf}). Thus, the formula (\ref{var-vf}) shows that the pre-symplectic 1-form on the space of solutions to field equations is given by
\be
\theta(\delta\w) = -\int_\Sigma {\rm Tr}\left(\delta \w \wedge \e_f\right).
\ee
Thus, a multiple of the spatial projection of $\e_f$ is the canonically conjugate momentum to $\w$. As we have seen from (\ref{e-a}), the spatial projection $\e_a$ is a function of $\dot{\w}_a-\nabla_a\w_0$ and the spatial curvature $\f_{ab}$. So, the momentum conjugate to $\w_a$ is a function of the time derivative of the connection, as is appropriate for a second order theory. Thus, when quantising the theory, a possible polarisation is to keep the connection on $\Sigma$ fixed and have wave functions depend on $\w|_\Sigma$. 

\subsection{Generalisation to gravity coupled to the 2-form field}

The action of $\Lambda=-1$ gravity to the 2-form field is just the sum of (\ref{s-matrix}) and (\ref{s-bc}). We have
\be
S[\e,\w,b] = -\int_M {\rm Tr}\left( \e\wedge \f - \frac{1-\chi^2}{3} \e\wedge \e\wedge \e\right).
\ee
Its variation with respect to $\e$ was computed in (\ref{f-e-chi}). It is clear that the solution of this equation can be written as 
\be\label{e-chi}
\e = \frac{\e_\f}{\sqrt{1+\chi^2}},
\ee
where $\e_\f$ solves $\f=\e_\f\wedge\e_\f$. Substituting this back into the action, after some simple algebra, we get the following connection plus 2-form field action
\be\label{s-wB}
S[\w,b] =-\frac{2}{3} \int_M \frac{1+2\chi^2}{(1+\chi^2)^{3/2}} {\rm Tr}\left( \e_\f \wedge \e_\f\wedge \e_\f\right).
\ee
Here $\chi$ is the solution of the following equation
\be\label{chi-B}
\frac{\chi}{(1+\chi^2)^{3/2}} = \frac{db+\omega}{v_\f},
\ee
where $v_\f$ is the volume form given by (\ref{volume-f}) and $\omega$ is the 3-form that introduces the coupling to the third cohomology of $M$. We see that the connection formulation action is still given by the volume of the space, but with a non-trivial pre-factor depending on the 2-form field (and $\omega$). 

\subsection{Field equations directly from the pure connection action}

It is instructive to see how the field equations, which are already familiar to us, arise directly from (\ref{s-wB}). First, let us see the field equation that says $\chi=const$. This is obtained by varying with respect to $b$. One first differentiates the $\chi$-dependent pre factor in (\ref{s-wB}). Then, to find the variation of $\chi$ one takes the variation of both sides of (\ref{chi-B}). It is then clear that the variation of the action with respect to $b$ is integral of a certain function of $\chi$ times $d(\delta b)$. Integrating by parts gives the equation $d\chi=0$. 

We can also consider the variation of the action as the connection $\w$ gets varied. To compute this one needs to consider both the variation of $\e_\f$, as well as the variation of $\chi$, because $\chi$ depends on the volume form $v_\f$ that in turn depends on the connection. However, it is clear that the result is proportional to the integral of some function of $\chi$ times the variation of $v_\f$. We have computed this variation above in (\ref{var-vf}). Together with the previously established equation $\chi=const$ this shows that (\ref{pure-conn-feqs}) results. This establishes that (\ref{s-wB}) is the correct pure connection description of 3D gravity coupled to the 2-form field. 

\subsection{The effective value of the cosmological constant}

We have defined the frame field $\e_\f$ via $\f=\e_\f\wedge \e_\f$. As we have just seen, the field equation for the connection says $\nabla\e_\f=0$, and thus the connection is metric compatible one. This means that the metric constructed from $\e_\f$ is of constant curvature corresponding to the value $\Lambda=-1$. 

However, in the pure connection formulation, once we integrated out the field $\e$, it is not clear whether it is the metric defined by $\e_\f$, or some constant multiple of it that should be considered as the "physical" metric. In the absence of coupling to any other matter that might be present, it is impossible to decide on this conclusively. However, there is one mathematical argument in favour of calling a certain metric as physical. Indeed, for $\chi\not=0$, the action (\ref{s-wB}) is not the volume of the metric determined by $\e_\f$, but rather the volume computed using certain rescaled frame field. Thus, it is clear that the "physical" frame field whose volume form is integrated in (\ref{s-wB}) is
\be
\e_{phys} = \frac{(1+2\chi^2)^{1/3}}{\sqrt{1+\chi^2}} \e_\f.
\ee
Note that this is different from (\ref{e-chi}). Thus, the mathematically natural normalisation of the frame does not coincide (when 2-form field is present) with the original normalisation in the first order action. 

\section{Lift from 3D to 6D}
\label{sec:lift}

The content of the previous sections was mostly a review of known facts, with some original elements in the description of the pure connection formulation of 3D gravity. We now turn to original part of this paper.

We first describe an intimate relationship between the 6D story of section \ref{sec:hitchin} and the pure connection formulation of section \ref{sec:pure}. As we shall see, the 6D interpretation is the best way to understand what the pure connection formulation is, and why it is possible. We consider pure gravity in this section, so both the 2-form field $b$ and 3-form $\omega$ are set to zero. 

\subsection{Connections on the principal ${\rm SU}(2)$ bundle}

This subsection is just to establish our bundle notations. Thus, we want to establish a relation between connections $\w$ as one-forms on $M$ with values in $2\times 2$ anti-hermitian matrices and a more geometrical description in terms of Lie algebra valued one-forms on the total space of the bundle. 

Consider the total space $\M$ of the principal ${\rm SU}(2)$ bundle over $M$, with copies of the group manifold attached over every base point. We choose a trivialisation of this bundle so that every fiber is identified with a copy of the group. We choose the group action to be the action of the group on itself by {\it right} multiplication. We then introduce the Maurer-Cartan form
\be
\m:= g^{-1} dg, \qquad g\in {\rm SU}(2).
\ee
We form the connection in the principal bundle
\be\label{trivialis}
\W:= g^{-1} \w g + \m.
\ee
This is a Lie-algebra valued 1-form in the total space of the bundle, whose kernel defines the notion of horizontal vector fields. An easy calculations gives the curvature of this connection 
\be
\F= d\W+\W\wedge \W = g^{-1} (d\w+\w\wedge \w) g \equiv g^{-1} \f g,
\ee
which is a horizontal 2-form. While both $\m,\w$ depend on the trivialisation chosen, the connection $\W$ is a geometrically well-defined object - a Lie-algebra valued 1-form on $\M$. Changes of trivialisation correspond to gauge transformations of $\w$. 

\subsection{Almost-complex structure on the principal ${\rm SU}(2)$ bundle}

To make a link to the 6D story of section \ref{sec:hitchin}, let us assume that $\M$ is the total space of the principal ${\rm SU}(2)$ bundle ${\rm SU}(2) \hookrightarrow P\to M$. Let $J$ be an ${\rm SU}(2)$ invariant almost-complex structure on $P$. The following proposition is known to experts, but we were unable to find a reference.
\begin{proposition} Let $J$ be an ${\rm SU}(2)$-invariant almost-complex structure on $\M\to M$ that does not have a vertical invariant subspace. Then $J$ defines an ${\rm SU}(2)$ connection and a metric on $M$. In turn, given a ${\rm SU}(2)$ connection and a metric on $M$, there is $J$ on $\M$ that returns this connection and the metric by the first part of this proposition. 
\end{proposition}
A proof in one direction is easy. Our assumption that there is no vertical invariant subspace is equivalent to the assumption that the image of vertical vector fields is no longer vertical. We can then declare the horizontal vector fields to be the $J$-image of vertical ones. This defines the notion of horizontality, and thus an ${\rm SU}(2)$ connection. Thus, by definition
\be
J(V)=H,
\ee
and we have a split $T\M=V\oplus H$. To define the metric, we take two horizontal vector fields $\xi_H,\eta_H\in H$, act on them with $J$ to obtain two vertical vector fields, and then pair the result using the Killing-Cartan metric on the Lie algebra:
\be
(\xi,\eta)_M:= (J_\Omega(\xi_H), J_\Omega (\eta_H))_{{\rm SU}(2)}.
\ee

A proof in the other direction is by explicit construction of $J$. Thus, let us take a connection on $M$ and lift it to the total space of the bundle, where it is represented by the Lie algebra-valued 1-form $\W$ on $\M$. It is convenient to introduce also the component one-forms $\W^i$ by writing $\W=\W^i \tau^i$, where $\tau^i$ are multiples of Pauli matrices introduced in (\ref{tau}). Then, a metric on the base can be lifted to an ${\rm SU}(2)$-invariant pairing of horizontal vector fields. This pairing can be written as $-2{\rm Tr}(\E\otimes\E)$ for some Lie algebra valued 1-form $\E$, which we also decompose as $\E=\E^i\tau^i$. The right ${\rm SU}(2)$ action on the fibers becomes the ${\rm SO}(3)$ action on $\W^i,\E^i$. We now form the dual basis of vector fields on $T\M$ that is determined from the following conditions
\be\label{dual-basis}
\W^j\left(\frac{\partial}{\partial \W^i}\right)  = \delta_i^j, \qquad \E^j \left(\frac{\partial}{\partial \W^i} \right) =0,
\\ \nonumber
\W^j\left( \frac{\partial}{\partial \E^i}\right)  =0, \qquad \E^j\left( \frac{\partial}{\partial \E^i}\right)  =\delta_i^j,
\ee
where the meaning of the left-hand-sides here is that the vector fields are inserted into the corresponding one-forms. Finally, we define the following almost-complex structure:
\be\label{J-EW}
J_{\E,\W}:= \E^i \otimes \frac{\partial}{\partial \W^i}  - \W^i \otimes \frac{\partial}{\partial \E^i}.
\ee
This almost-complex structure is clearly ${\rm SO}(3)\sim {\rm SU}(2)$ invariant. By construction, this sends vertical vector fields $\partial/\partial \W^i$ to horizontal $\partial/\partial \E^i$, and vice versa. It is also clear that notion of the connection that is defined by the above $J_{\E,\W}$ coincides with $\W$ and that the metric $J_{\E,\W}$ defines is that for the frame $\E$. $\qed$

\begin{corollary} The $(0,1)$ forms with respect to almost-complex structure (\ref{J-EW}) are given by 
\be\label{A}
A:=\W+\im\E.
\ee 
\end{corollary}

A proof is by rewriting the formula (\ref{J-EW}) in the following way
\be\label{J-hor-vert}
J_{\E,\W} = \frac{1}{\im} ( \W+ \im \E)^i \frac{1}{2} \left( \frac{\partial}{\partial \W} -\im \frac{\partial}{\partial \E}\right)_i -  \frac{1}{\im} ( \W- \im \E)^i \frac{1}{2} \left( \frac{\partial}{\partial \W} +\im \frac{\partial}{\partial \E}\right)_i. 
\ee
$\qed$

\begin{proposition} The almost-complex structure $J_{\E,\W}$ is integrable if and only if the ${\rm SL}(2,\C)$ connection $A$ in (\ref{A}) is flat $F(A)\equiv dA+A\wedge A=0$. Thus, $J_{\E,\W}$ is integrable if and only if the connection $\W$ is torsion free $\nabla\E=0$ and the metric is of constant negative curvature $F(\W)=\E\wedge\E$.
\end{proposition}

Flatness of $A$ is equivalent to integrability of $J_{\E,\W}$ in view of Frobenius theorem. The fact that flatness of $A$ is equivalent to $\nabla\E=0$ and $F(\W)=\E\E$ is a simple computation. $\qed$

\begin{corollary} The almost-complex structure $J_{\E,\W}$ is one in the sense of Hitchin for the 3-form $\Omega_{\E,\W}=2{\rm Re}(\Omega_{\E,\W}^c)$ with
\be
\Omega^c_{\E,\W}: = -\frac{1}{6} {\rm Tr} \left( A^3\right).
\ee
\end{corollary}

This is clear from (\ref{omega-c}). $\qed$

\subsection{Chern-Simons 3-form}

In Section \ref{sec:hitchin} we have seen that stable 3-forms (of negative type) define an almost-complex structure. In the previous subsection we have seen that when the 6D manifold has the structure of the total space of the principal ${\rm SU}(2)$ bundle, this in turn defines a connection and a frame. We now take a particular 3-form that is {\it closed} and parametrised solely by an ${\rm SU}(2)$ connection. We than check that the connection this 3-form defines is the connection we started from. But the 3-form also defines the metric. This metric turns out to be just the metric as defined by the connection as we have explained in section \ref{sec:pure}. This gives a 6D lift of the pure connection formulation of 3D gravity. To a large extent, this also gives an explanation for why the pure connection formulation exists. 

Thus, starting from an ${\rm SU}(2)$ connection $\W$ on $P$, we form the Chern-Simons 3-form 
\be\label{omega-CS}
\Omega = CS(\W)\equiv {\rm Tr}\left( \W\wedge d\W + \frac{2}{3} \W\wedge \W \wedge \W\right).
\ee
By construction, this is a well-defined right-invariant 3-form on $\M$. The exterior derivative of this 3-form is given by
\be\label{d-omega}
d\Omega = {\rm Tr}\left( d\W\wedge d\W + 2 d\W\wedge \W\wedge \W\right) = {\rm Tr}( \F\wedge \F)={\rm Tr}( \f\wedge \f)=0,
\ee
where we have used ${\rm Tr}(\W\wedge \W\wedge \W\wedge \W)=0$. This identity follows by expanding each $\W$ into the basis of Lie algebra generators, and noticing that the trace produces contractions of the type $\W^i\wedge \W^i=0$. The calculation in (\ref{d-omega}) shows that $d\Omega$ is a 4-form on the 3-dimensional base, and thus vanishes. Thus, the Chern-Simons 3-form (\ref{omega-CS}) on the total space of the bundle is closed. 

\subsection{Almost-complex structure for $CS(\W)$}

We now want to impose the condition that (\ref{omega-CS}) is of negative type, i.e. defines an almost-complex structure. To see what this implies for the connection $\W$ we first rewrite $\Omega$ in terms of the curvature
\be\label{omega-curvature}
\Omega ={\rm Tr}\left( - \frac{1}{3} \W\wedge \W\wedge \W + \W\wedge \F\right).
\ee
We want this to be equal to (twice) the real part of some complex decomposable 3-form $\Omega=2{\rm Re}(\Omega^c)$ with 
\be\label{omega-c}
\Omega^c = -\frac{1}{6} {\rm Tr}\left( A\wedge A\wedge A \right),
\ee
for an appropriate ${\rm SL}(2,\C)$-valued 1-form $A$. We know from the Section \ref{sec:pure} that there are two possible signs of $\F$ in (\ref{omega-curvature}). In the negative case, $\f=\e_\f \wedge \e_\f$ and we can lift this relation to the total space of the bundle as $\F=\E\E$ with $\E:=g^{-1} \e_\f g$. In the positive sign case, we similarly have $\F=-\E\E$. It is clear that the negative sign case corresponds to (\ref{omega-curvature}) being of negative type and defining the almost-complex structure with the holomorphic 3-form (\ref{omega-c}) and 
\be\label{A-WE}
A=\W+\im\E.
\ee
The positive sign case $\F=-\E\E$ renders (\ref{omega-curvature}) of positive type. There is no almost-complex structure in this case. The above discussion establishes equivalence of two possible definitions of negative definite connections, one given in the Introduction, and the other one of Section \ref{sec:pure}

\begin{proposition} The Chern-Simons 3-form $CS(\W)$ in the total space of the bundle is of negative type, i.e. defines an almost-complex structure, if and only if the connection $\W$ is negative definite so that $\F=\E\E$. 
\end{proposition}

Another way to see the equivalence of notions in Sections \ref{sec:hitchin} and \ref{sec:pure} is by noticing that the map $\phi_\f$ defined in (\ref{phi-def}) is just the application of the almost-complex structure $J_\W$ as defined by $\Omega=CS(\W)$ to horizontal one-forms. Indeed, we know that $J_\W$ maps horizontal one-forms to vertical ones, and the latter can be identified with elements of the Lie algebra. This is exactly what the map $\phi_\f$ does. 

We also see from (\ref{A-WE}), (\ref{omega-c}) and the discussion of the previous subsection that $\W$ is just the connection defined by the almost-complex structure defined by $CS(\W)$. This establishes part of the Proposition 1 of the Introduction. It is also clear that a frame for the metric defined by the almost-complex structure for $CS(\W)$ is $\E$. This establishes the other part of Proposition 1. 

\subsection{Calculation of the exterior derivative of $\Omega^c$}

We already know from Proposition 5 above that the almost-complex structure for $CS(\W)$ will be integrable iff $\nabla\E=0$. Let us spell out the corresponding computation. 

Field equations of the Hitchin theory can be combined into the statement that the complex 3-form $\Omega^c$ is closed. We have
\be
0= d\Omega^c = - \frac{1}{2} {\rm Tr}\left( dA\wedge A\wedge A\right) = - \frac{1}{2} {\rm Tr}\left( \F_A\wedge A\wedge A\right),
\ee
where we again have used the fact that the trace of the wedge product of 4 matrix-valued one-forms is zero.  The above equation implies
\be
0=\F_A = dA + A\wedge A = \F + \im \nabla\E - \E\wedge\E.
\ee
The real part of this equation is satisfied by construction of $\E$, while the imaginary part is the field equation (\ref{pure-conn-feqs}) of the pure connection formulation of 3D gravity. This proves Proposition 2 of the Introduction. 

\subsection{Calculation of the action}

We now compute the Hitchin's action (\ref{action-hitchin}) on our 3-form $\Omega$. We use the expression (\ref{v-omega}), and rewrite $\hat{\Omega}=2\,{\rm Im}(\Omega^c)$ in terms of $\W$ and $\E$. We have
\be\label{hat-omega}
\hat{\Omega} =  {\rm Tr}\left( \frac{1}{3} \E\wedge \E\wedge \E - \W\wedge \W\wedge \E)\right).
\ee
We also use the expression (\ref{omega-curvature}) in which we replace $\F=\E\wedge \E$. This gives
\be
v_\Omega = -\frac{1}{2} \left[ \frac{1}{3} {\rm Tr} (\W^3) \frac{1}{3} {\rm Tr}(\E^3) + {\rm Tr}\left( \W \E^2\right) {\rm Tr}\left( \W^2 \E\right)\right],
\ee
where we omitted the wedge product signs. We can now replace $\W$ here by the Maurer-Cartan form $\m=g^{-1} dg$, as the part of $\W$ that is a 1-form on the base does not contribute. We can also rewrite the last term here as a multiple of the first term, using some simple properties of the Lie algebra generators $\tau^i$
\be\label{trace-ident}
{\rm Tr}\left( \W \E^2\right) {\rm Tr}\left( \W^2 \E\right)= \frac{1}{3} {\rm Tr} (\W^3) {\rm Tr}(\E^3).
\ee
Thus, overall
\be\label{v-omega}
v_\Omega = - \frac{2}{9} {\rm Tr} (\m^3) {\rm Tr}(\e_\f^3) = \frac{1}{3} {\rm Tr} (\m^3) \, v_\f,
\ee
where the volume form on the base $v_\f$ is given by (\ref{volume-f}). Thus, modulo the volume of the fiber, which is just the volume of ${\rm SU}(2)$, we have that the Hitchin action is equal to the action of pure connection 3D gravity. This proves a part of the Proposition 3 of the Introduction. The part of this Proposition that states that the torsion free condition for $\w$ arises as the Euler-Lagrange equation following by extermising the volume for $\e_f$ was already proven in Section \ref{sec:pure}.

\subsection{Metric on the total space}

We note that not only we have a naturally defined notion of the metric on the base, but also there is a metric on the total space $\M$. Indeed, this comes from extending the metric on the fibers to a Hermitian metric on the total space
\be\label{metric-total-space}
ds^2_{\M} := \delta_{ij} A^i \overline{A^j},
\ee
where $A^i$ are the holomorphic 1-forms. This metric, together with the complex structure give rise to the 2-form 
\be
w: = \frac{\im}{2} \delta_{ij} A^i \wedge \overline{A^j} = \W^i \wedge \E_i.
\ee
However, as it is easy to check, this form is not closed, and so the metric arising is not K\"ahler. 

\section{Dimensional reduction from 6D to 3D}
\label{sec:reduction}

We now consider the more complicated issue of dimensional reduction. One of the main purposes of this section is to see how the ansatz (\ref{omega-CS}) that we have used in the previous section arises naturally. We will also encounter some subtleties as to interpretation of Hitchin's theory in the context of dimensional reduction. 

\subsection{Assumption of invariance}

To analyse the dimensional reduction we need to assume that fields of some theory in a bigger number of dimensions do not depend on some of the coordinates so that they can be interpreted as fields in a smaller number of dimensions. A geometrical way of doing this is to say that we have some group acting in $\M$, and the fields on $\M$ are invariant under this action. 

Thus, we shall assume that we have a free and transitive action of the group ${\rm SU}(2)$ on $\M$. The space of orbits is a manifold, which we denote by $M$. This realises $\M$ as a principal ${\rm SU}(2)$ bundle over $M$. Because $\M$ is 6-dimensional, $M$ is 3-dimensional, and ${\rm SU}(2)$ bundles over a 3-dimensional base are trivial. So, $\M$ has the structure of the product manifold $\M=M\times {\rm SU}(2)$. 

\subsection{Closed ${\rm SU}(2)$ invariant 3-forms}

The purpose of several subsections that follow is to prove the following proposition
\begin{proposition} Any closed ${\rm SU}(2)$ invariant 3-form in the total space of the bundle $\M\to M$ can be written as
\be\label{omega**}
\Omega= CS(\W)  + \omega + db
\ee
for some connection $\W$ on $\M$, a representative $\omega, d\omega=0$ of $H^3(M)$, and a 2-form field $b\in \Lambda^2(M)$ on the base. 
\end{proposition}
Once this fact is established, we perform a computation to understand what the Hitchin field equation $d\hat{\Omega}=0$ implies about the fields $\W,\omega,b$. The reader that is not interested in details of this proof can skip directly to subsection 7.7, where the field equation is analysed. There is also a subtlety which is that the form (\ref{omega**}) does not stay in the same cohomology class if one varies the connection $\W$. This subtlety is discussed in subsection 7.6. 

\subsection{Specifying the cohomology class}

To prove the above proposition, we start by fixing the cohomology class of $\Omega$. Thus, let $\Omega_*$ be a fixed closed 3-form on the total space $\M$. To describe this 3-form concretely, we use the classical result of K\"unneth, see e.g. \cite{Bott}, page 47, that says that the cohomology of the product of manifolds is the product of cohomologies. Concretely, in our case this means that the third cohomology of $P$ is generated by the third cohomology of the fiber times scalars on the base and third cohomology of the base time scalars on the fiber. 

The complication arises from the fact that, even though the bundle $P\to M$ is trivial, there is no canonical trivialisation. Thus, let $s:M\to P$ be a section. Using this section we can identify each fiber with a copy of the group. This defines a projection $P\to {\rm SU}(2)$, but this projection depends on the section $s$. We can now pull back the cohomology from the base using the projection $P\to M$, and pull back the cohomology from ${\rm SU}(2)$ by using the projection defined by the section. But since the projection $P\to {\rm SU}(2)$ depends on a section, there is no canonical way of describing cohomology of the total space as the product of cohomologies. This is the source of subtleties that we will need to address in this section. 

Let us describe all this in coordinates, which is what is convenient for practical computations. Choosing a section $s$ we identify fibers with copies of the group. Adapted to this identification are coordinates $(g,x)$ on $P$. K\"unneth theorem guarantees that every representative of $H^3(P)$ can be written as 
\be\label{omega*-1}
\Omega_* = -\frac{c_1}{3} {\rm Tr}(\m^3) + c_2\, \omega,
\ee
where as before $\m=g^{-1} dg$ is the Maurer-Cartan form on the group, $\omega\in H^3(M)$ is a closed 3-form on the base, and $c_1,c_2$ are constants. Since the Hitchin action is homogeneous of degree 2 in $\Omega$, if $c_1\not=0$, which is what we shall assume, we can always rescale $\Omega_*$ by changing the coefficient in front of the Hitchin action to achieve $|c_1|=1$ in (\ref{omega*-1}). We can then use the circle action discussed in Section \ref{sec:hitchin} to set $c_1=1$. This changes also $c_2$, but this constant can always be absorbed into $\omega$. 

We can now see the subtlety that arises because $\Omega_*$ above depends on the section $s:M\to P$ used to trivialise the bundle. Indeed, choices of trivialisations are related by "gauge transformations"
\be
\m = g^{-1} dg \to g^{-1} k^{-1} d(k g) = g^{-1} dg + g^{-1} (k^{-1} dk) g,
\ee
where $k: M\to {\rm SU}(2)$ is a group valued field on the base. It is easy to check that under this transformation
\be\label{mc-change}
-\frac{1}{3}{\rm Tr}(\m^3) \to -\frac{1}{3}{\rm Tr}(\m^3) - d\left({\rm Tr}( \m g^{-1} (k^{-1} dk) g)\right) - \frac{1}{3}{\rm Tr}((k^{-1} dk)^3).
\ee
The second term here is exact, but the third term, which is purely basic, is in general not exact. In the case of $M=S^3$, its integral over the base vanishes for "small" gauge transformations, but in general measures the winding number of the transformation. We will return to this subtlety below, after we discuss the exact terms that can be added to (\ref{omega*-1}). 

\subsection{2-forms $B$ modulo 1-forms}

With the choice of normalisation as discussed above, the general 3-form in the cohomology class of (\ref{omega*-1}) is given by
\be\label{omega-dim-red}
\Omega = -\frac{1}{3} {\rm Tr}(\m^3) + \omega + dB, \qquad B\in \Lambda^2(\M).
\ee
We assume that $B$ in (\ref{omega-dim-red}) is a $G$-invariant 2-form on $\M$, and write a general decomposition of such a form using the 1-forms $\m$ as the basis of 1-forms for the fiber directions. We have
\be\label{H-dim-red}
B= {\rm Tr}(g^{-1} \phi g \m \m) + {\rm Tr}(H \m) + b.
\ee
Here $\phi$ is a ${\mathfrak su}(2)$-valued function on $M$, $H=g^{-1} \h g$ with $\h$ a ${\mathfrak su}(2)$-valued 1-form on the base, and $b \in \Lambda^2(M)$. It is clear that $\phi,\h,b$ capture all the components of a general 2-form on $\M$. Indeed, the dimension of $\Lambda^2(\M)$ is $15$, and we have 3 functions in $\phi$, $3\times 3=9$ functions in $\h$ and $3={\rm dim}(\Lambda^2(M))$ functions in $b$. 

We now use the fact that $B$ is defined modulo $dT$, where $T\in \Lambda^1(\M)$ is in turn defined only modulo the differential of a function. We again assume that $T$ is $G$-invariant, and write its decomposition
\be
T = {\rm Tr}(g^{-1} \psi g \m) + \eta, \qquad \eta \in \Lambda^1(M).
\ee
Computing the exterior derivative of $T$ we get
\be
dT = {\rm Tr}\left(g^{-1} d\psi g\, \m  + g^{-1} \psi g \m \m \right) + d\eta.
\ee
We thus see that by choosing the Lie-algebra valued functions $\psi$ appropriately, we can cancel the first term in (\ref{H-dim-red}), as $\psi$ shifts $\phi$ (and also shifts $h$ by $d\psi$). It is very convenient to do so as this will simplify computations dramatically. So, we now fix (a part of) our ambiguity in choosing $B$ by taking it to be of the form
\be\label{H-fixed}
B=  {\rm Tr}(H \m) + b/d\eta,
\ee
where the last term stands for $b\in \Lambda^2(M)$ modulo $d\eta, \eta\in \Lambda^1(M)$, where in turn $\eta$ is only defined modulo the differential of a function. Overall, we have $3\times 3$ functions on $M$ left in $\h$, and $3-(3-1)=1$ function left in $b/d\eta$.

\subsection{Interpretation of $H$ as a connection}

It is now a matter of a simple calculation to show that
\be\label{omega*}
\Omega= CS(\W)  + \omega + db -{\rm Tr}\left( \h d\h + \frac{2}{3}\h^3\right),
\ee
where we introduced a new Lie-algebra valued one-form
\be
\W = \m + H,
\ee
and $CS(\W)$ is the Chern-Simons 3-form (\ref{omega-CS}). 

We cannot yet identify $\W$ with a connection in $\M$ because we did not specify how $H$ transforms under changes of the trivialisation. Let us declare that it transforms so that $\W$ is a geometric object that does not depend on the trivialisation used to define it. Thus, we want $\W$ to be a connection in $\M$. With this transformation rule $\h$ transforms as a connection field on the base
\be
\h \to k^{-1} \h k + k^{-1} dk, \qquad g\to kg.
\ee
With this transformation rule we have
\be\label{cs-change}
{\rm Tr}\left( \h d\h + \frac{2}{3}\h^3\right) \to {\rm Tr}\left( \h d\h + \frac{2}{3}\h^3\right) - \frac{1}{3}{\rm Tr}((k^{-1} dk)^3) - d\left({\rm Tr}( dk k^{-1} \h )\right).
\ee
The second non-exact term here is the same term that already appeared in (\ref{mc-change}). 

\subsection{Closed 3-forms vs. 3-forms in the same cohomology class}

The formula (\ref{omega*}) was established by considering a fixed representative of the cohomology $H^3(\M)$ and adding to it an invariant exact form. Noticing that the last term in (\ref{omega*}) is purely basic, we see that it can be combined with $\omega$. This establishes our Proposition 7 above. 

We now come to the subtlety that arises in interpreting Hitchin theory in the context of dimensional reduction. This theory is about finding a representative of a fixed cohomology class that makes the associated almost complex structure integrable. Once found, this representative will be a closed invariant 3-form $\Omega$ satisfying the equation $d\hat{\Omega}=0$. Since we established that any closed invariant 3-form can be written as (\ref{omega**}), we can proceed with analysing the implications of the equation $d\hat{\Omega}=0$ for the data that appear in (\ref{omega**}), and this will be done in the following subsections. 

The subtlety is that while any closed invariant 3-form can be written as (\ref{omega**}), forms in the same cohomology class are not parametrised by (\ref{omega**}) with varying $\W,b$. We have seen this very clearly from (\ref{omega*}) that provides such a parametrisation of forms in a fixed cohomology class, but at the expense on depending on a fixed section of the bundle. 

This can be further clarified as follows. In Hitchin's theory we search for a form $\Omega$ in a fixed cohomology class, i.e. we fix a closed 3-form $\Omega_*$ and consider $\Omega=\Omega_* + dB$ for some invariant 2-form $B$. Varying the action with respect to $B$ we get the equation $d\hat{\Omega}=0$. However, to obtain this parametrisation of the cohomology class we had to make a choice of $\Omega_*$. This choice introduces some "background" structure. In particular, the 3-form $\Omega_*$ itself defines a connection on $\M$, from the construction that we have explained in the previous section. In the concrete parametrisation that we introduced above and that led to (\ref{omega*}) this background structure was that of a fixed section of the bundle, and the connection defined by $\Omega_*$ was the associated flat connection. It is clear that in this parametrisation, which is important for obtaining the correct field equations, the 3-form $\Omega$ is specified by the choice of the background structure $\Omega_*$ as well as by the deviation $dB$ from it. This parametrisation ignores the fact that the form $\Omega$ itself defines a geometric structure. It is the parametrisation (\ref{omega**}) that describes $\Omega$ in terms of the geometric structures that it defines. It is thus the parametrisation (\ref{omega**}) that is appropriate for analysing the geometric consequences of the equation $d\hat{\Omega}=0$. 

The subtlety is then that we must use the parametrisation (\ref{omega*}) of forms in a fixed cohomology class to establish the field equations, but parametrisation (\ref{omega**}) to analyse their consequences. If we instead take take the Hitchin action and evaluate it on the ansatz (\ref{omega**}) and then vary with respect to $\W,\omega$ we obtain some other equation instead of $d\hat{\Omega}=0$. This other equation is also potentially interesting, as we explain at the end of this section. But for now we proceed with unravelling the consequences of $d\hat{\Omega}=0$.

\subsection{Calculation of $\hat{\Omega}$}

To see what the Hitchin's field equations imply for fields $\w,b$ on the base, we need to compute the 3-form $\hat{\Omega}$. To do this, as in the previous section, we assume that the curvature $\f$ of $\w$ solves the equation $\f=\e_\f\wedge \e_\f$. This is the assumption that the connection $\w$ is negative definite. We can then use the set of 1-forms $\e_\f$ to introduce the volume form on the base ${\rm Tr}(\e_\f^3)$. We then write the last two terms in (\ref{omega**}) as a multiple of the volume form on the base. This gives
\be\label{omega-rho}
\Omega = {\rm Tr}\left( - \frac{1}{3} \W^3 + \E \E \W \right) -\frac{2\rho}{3} \,{\rm Tr}( \E^3).
\ee
Here 
\be
\rho := \frac{\omega+db}{v_\f}
\ee
is some function on the base, and $\E \equiv g^{-1} \e_\f g$ is the lift to the total space. 

\subsection{Direct computation}

One way to compute $\hat{\Omega}$ is to write $\Omega = 2{\rm Re}(\Omega^c)$ with 
\be
\Omega^c = -\frac{1}{6} {\rm Tr}\left( \cA^3 \right)
\ee
with $\cA$ of the form
\be\label{A-twisted}
\cA = \alpha \W + \beta \E + \im ( \gamma \W + \delta \E ),
\ee 
for some functions $\alpha,\beta,\gamma,\delta$. We get the following equations for $\alpha,\beta,\gamma,\delta$
\be\label{eqs}
- \alpha^2 \beta+ \gamma^2 \beta + 2\alpha \gamma\delta=0, \quad -\frac{1}{3} \alpha^3 + \alpha \gamma^2 =-\frac{1}{3} \\ \nonumber
- \alpha \beta^2 + \alpha \delta^2 + 2\beta \gamma\delta = 1, \quad -\frac{1}{3} \beta^3 + \beta \delta^2 = -\frac{2}{3} \rho.
\ee
This is sufficient to find $\alpha,\beta,\gamma,\delta$ in terms of $\rho$. We present an explicit solution to these in the Appendix. One can then analyse what the equation $d\hat{\Omega}=0$ implies. This computation is straightforward, but somewhat involved. Again, we give details in the Appendix. We just quote the result here, and present a shortcut in the next subsection. The result is that 
\be\label{E-eqn}
d\hat{\Omega}=0 \Leftrightarrow d\rho=0, \qquad \nabla\E = 2\rho \,\,\E\E.
\ee
The first of these equations says that $\rho=const$, which is as expected. When $\rho=0$ the second equation says $\nabla\E=0$ and the connection is metric compatible. However, we see that in general the situation is more complicated and the connection is not torsion free.

\subsection{Circle action and the twist}

There is a much quicker way of obtaining the results quoted above. This way uses a trick with the circle action in the space of 3-forms. Consider the following complexified Lie-algebra valued 1-form
\be
\bar{A}:= \W+ \im e^{\im\theta} \E.
\ee
Here $\theta$ is at this stage arbitrary angle variable. When $\theta=0$ we have the Chern-Simons connection considered in the previous section. For non-zero $\theta$ there is a $\E$ proportional contribution to the real part of $\bar{A}$. We can consider 3-forms ${\rm Re}({\rm Tr}(\bar{A}^3))$ and ${\rm Im}({\rm Tr}(\bar{A}^3))$. Simple inspection reveals that the object
\be
\bar{\Omega} := -\frac{1}{3} \left( \cos(\theta) {\rm Re}({\rm Tr}(\bar{A}^3))+\sin(\theta) {\rm Im}({\rm Tr}(\bar{A}^3))\right)
\ee
is of the form (\ref{omega-rho}) in the sense that there is no term $\W^2\E$. A simple computation gives
\be
\bar{\Omega} = \cos(\theta){\rm Tr} \left( - \frac{1}{3} \W^3 + \E \E \W -\frac{2\sin(\theta)}{3}  \E^3\right).
\ee
In other words, if we identify
\be
\rho= \sin(\theta)
\ee
we get
\be
\bar{\Omega} = \cos(\theta) \Omega.
\ee
It is clear that the 3-form $\bar{\Omega}$ admits the following simple representation
\be
\bar{\Omega} = 2{\rm Re} \left( e^{-\im\theta} \bar{\Omega}^c\right)
\ee
with 
\be\label{bar-omega-c}
\bar{\Omega}^c := -\frac{1}{6} {\rm Tr}(\bar{A}^3).
\ee
Therefore, we can write
\be\label{omega-rep}
\Omega = \frac{1}{\cos(\theta)} 2{\rm Re} \left( e^{-\im\theta} \bar{\Omega}^c\right).
\ee
This immediately allows us to read off the complex 3-form whose real part is $\Omega$. 

The equations $d\Omega=0, d\hat{\Omega}=0$ are then equivalent to the statement that $\bar{\Omega}^c$ is closed. This, in turn, is equivalent to the condition that the "connection" $\bar{A}$ is flat, together with the condition $d\rho=0$. The flatness equation reads
\be
0=\F(\bar{A}) = \F(\W) + \im e^{\im\theta} \nabla \E - e^{2\im\theta} \E\E.
\ee
The imaginary part of this equation is
\be\label{E-eqn-cos}
\nabla\E-2\sin(\theta)\E\E=0,
\ee
and coincides with (\ref{E-eqn}). The real part of this equation is
\be
\F - \sin(\theta) \nabla\E - (\cos^2(\theta)-\sin^2(\theta)) \E\E=0.
\ee
Taking into account that $\F=\E\E$ by definition of $\E$, we see that this coincides with (\ref{E-eqn-cos}), and is thus satisfied provided (\ref{E-eqn-cos}) is satisfied. 

\subsection{Interpretation}

The analysis of the previous subsection tells us that the complex structure determined by 3-form (\ref{omega-rho}) is such that the 1-forms $\bar{A}$ are holomorphic. Writing this "connection" in terms of the original parameter $\rho$ we have
\be\label{bar-A}
\bar{A} = \W -\rho \E + \im \sqrt{1-\rho^2} \E.
\ee
This maps the current problem into what we have considered in the previous section. Indeed, the almost-complex structure with (\ref{bar-omega-c}) as the holomoprhic 3-form defines a connection and a metric. The connection is ${\rm Re}(\bar{A})$ and the frame field is ${\rm Im}(\bar{A})$. Thus, we see that the geometrically natural connection as defined by the complex structure defined by (\ref{omega-rho}) is
\be\label{wJ}
\W_J := \W-\rho\E.
\ee
We also confirm this expression for the geometric connection by a direct calculation using (\ref{A-twisted}) in the Appendix. 
The equation (\ref{E-eqn}), which, as we know, is equivalent to the flatness of $\bar{A}$, is then simply the statement that the connection $\W_J$ is compatible with $\E$
\be
\nabla_J \E=0.
\ee
This identifies $\rho\E$ as the torsion term in $\W=\W_J+\rho\E$, with $\W_J$ being metric compatible. 

The metric compatibility of $\W_J$ is the imaginary part of the flatness of $\bar{A}$ condition. The real part of this condition becomes the Einstein equation for the metric connection
\be
\F_J = (1-\rho^2)\E\E.
\ee
All in all, we have recovered all the ingredients of the pure connection formulation of the 3D gravity coupled to the topological 2-form field. Indeed, we see that the field strength $db$ must be constant in the sense that $(\omega+db)/v=\rho$, and that the metric must be of constant curvature, with the effective value of the cosmological constant determined by $\rho$.

\subsection{Calculation of the action}

It is also possible to compute the value of the Hitchin action on the ansatz (\ref{omega**}), and then see what the resulting Euler-Lagrange equations imply. The easiest way to compute the action is to use the representation (\ref{omega-rep}), and appeal to the results of the previous section where we already computed the Hitchin action. It is clear that the result in the present case is some multiple of the volume form for $\E$. To compute the $\rho$-dependent coefficient we need to trace the factors of $\rho$. Because the action is homogeneity two in $\Omega$, there is a factor of $1/\cos^2(\theta)$ coming from the factor $1/\cos(\theta)$ in (\ref{omega-rep}). There is another factor of $\cos^3(\theta)$ coming from the fact that the action gives the volume of the metric with frame $\sqrt{1-\rho^2} \E$. Overall, using (\ref{v-omega}) we get
\be\label{v-wb}
v_\Omega = \frac{1}{3} {\rm Tr}(\m^3) \sqrt{1-\rho^2} \, v_\f.
\ee
Now, computing the variation of this action with respect to $b$ one will obtain the condition $d\rho=0$, which we know is the correct equation for the two-form field. We can also compute the variation with respect to the connection. Similar to what we already observed in section \ref{sec:pure}, the variation will give rise to the equation saying that $\w$ is metric compatible, i.e. that the torsion is zero. This is in contrast to the conclusion that follows from imposing $d\hat{\Omega}=0$, which implies that $\w$ has torsion. 

Thus, if we use the parametrisation (\ref{omega**}) and vary the Hitchin action within this class of 3-forms, because now $\Omega$ does not stay in the same cohomology class as one varies the connection, one no longer has the closeness of $\hat{\Omega}$ arising as the field equation. As we saw, the equation that arises is that $\w$ is torsion free, and then the curvature of $\w$ is constant. One then finds $d\hat{\Omega}$ to be a constant multiple of ${\rm Tr}(\E^2\W^2)$ and thus non-zero. So, with this interpretation the arising Euler-Lagrange equations no longer say that the almost-complex structure is integrable. But in terms of 3D fields one still obtains equations of gravity coupled to the topological two-form field. 

So, there is another possible interpretation of the Hitchin theory, at least in the present context of the dimensional reduction, which is to take the Hitchin action on 3-forms parametrised as in (\ref{omega**}). The arising 3D field equations are still sensible, but the 6D almost complex structure is no longer integrable. It is interesting that non-integrable almost complex structures naturally appear within a construction that embeds 6D manifolds as hyper surfaces in 7D.  We now turn to a description of this. 

\section{7D Interpretation}
\label{sec:7d}

This section is mainly for completeness purposes, and can be skipped at first reading. As we have already mentioned in the Introduction, there is also a 7D lift of the 3D structures described in \cite{BS}. This reference explains how 7D metrics of $G_2$ holonomy can be realised in the total space of a $\C^2$ spinor bundle over a 3-dimensional base. The most important fact about geometry in seven dimensions is that the tangent space at every point, viewed as a copy of $\R^7$, can be identified with the space of imaginary octonions. One then gets the operation of cross-product $\R^7 \times \R^7 \to \R^7$. This operation can be used for the following two constructions. First, given the usual flat metric $g(\cdot,\cdot)$ in $\R^7$ the cross-product associates with it the canonical 3-form $\Omega(X,Y,Z)=g(X\times Y, Z)$. In turn, a generic 3-form in 7D defines a metric. Second, given a codimension one hyper surface in a 7D manifold  one can take its normal vector $N$ (of unit norm) and define an almost complex structure on vectors tangent to the hyper surface via $J(X):=N\times X$. This almost complex structure is, in general, non-integrable, see \cite{Calabi}. 

The purpose of this section is two-fold. First, we show how a 7D construction explains where the metric (\ref{metric-total-space}) on the total space of the bundle comes from. Second, we make remarks on why it is natural to obtain non-integrable almost complex structures if 6D manifold is viewed as embedded into 7D. 

\subsection{3-Forms in 7D}

Here we review very briefly the geometry of 3-forms in 7 dimensions. For more details the reader is referred to \cite{Herfray:2016azk} and references therein. The basic fact about a generic 3-form is that it can be written in a canonical form
\be\label{omega-7-can}
\Omega = e^5\wedge e^6\wedge e^7+ e^5\wedge \Sigma^1 + e^6\wedge \Sigma^2 + e^7\wedge \Sigma^3,
\ee
where
\be
\Sigma^1 = e^1 \wedge e^2 - e^3\wedge e^4, \quad  \Sigma^2 = e^1 \wedge e^3 - e^4\wedge e^2, \quad  \Sigma^3 = e^1 \wedge e^4 - e^2\wedge e^3,
\ee
where $e^I, I=1,\ldots,7$ are 1-forms. Once such a canonical form is achieved, the metric defined by $\Omega$ is just the metric with $e^I$ as the frame
\be
ds^2 = \sum_I e^I\otimes e^I.
\ee

\subsection{From 6D to 7D}

We would like to construct a stable 3-form in 7D that is parametrised by a connection. There are various options to try, and some of them are preferred because they can also give rise to $G_2$ holonomy metrics. In this subsection we will not be concerned with the properties of the 7D metric, and just want to sketch the simplest possible construction that gives rise to the 6D metric we have encountered above. 

We consider
\be\label{omega-7}
\Omega :=  f(t)^3 CS(\W) + dt\wedge {\rm Tr}\left( \W\wedge \E\right)
\ee
on the space $\M\times \R$. In view of $\C^2=\R^4=S^3\times\R$ this can be interpreted as a 3-form in a $\C^2$ vector bundle over the 3D base. Thus, one may wonder if this is the form that appears in 
\cite{BS} in the construction of the metrics of holonomy $G_2$. The answer is negative, \cite{BS} considers a similarly looking form, but with $CS(\W)$ in the first term replaced by a form of the type $\E^3+\E\W^2$, see next subsection. In contrast to the construction in \cite{BS}, the form (\ref{omega-7}) is not closed, because the exterior derivative of ${\rm Tr}\left( \W\wedge \E\right)$ in the second term is not cancelled by $CS(\W)$. In spite of the fact that the exterior derivative of (\ref{omega-7}) is not zero, we proceed with the computation of the corresponding 7D metric. 

Let us rewrite (\ref{omega-7}) in the canonical form. We write $\W=\W^i \tau^i, \E=\E^i \tau^i$ where $\tau^i$ are the generators (\ref{tau}), and rewrite (\ref{omega-7}) as follows
\begin{align}
& \Omega = f^3 \W^1\wedge \W^2\wedge \W^3 + \W^1\wedge \left( dt\wedge \E^1 - f^3 \E^2\wedge\E^3\right) \\ \nonumber
& + \W^2\wedge \left( dt\wedge \E^2 - f^3 \E^3\wedge\E^1\right) 
+ \W^3\wedge \left( dt\wedge \E^3 - f^3 \E^1\wedge\E^2\right).
\end{align}
It is now clear that we can identify
\be
e^{4+i} = f\W^i, \quad e^{1+i}= f\E^i, \quad e^1 = f^{-2} dt,
\ee
after which the form (\ref{omega-7}) becomes (\ref{omega-7-can}). This gives for the metric
\be
ds^2 = f^{-4} dt^2 + f^2\sum_i \left( \W^i\otimes \W^i + \E^i\otimes \E^i\right).
\ee
The 6D restriction of this metric is a multiple of (\ref{metric-total-space}).  

We thus get another explanation of where the 6D metric (\ref{metric-total-space}) comes from. An ${\rm SU}(2)$ connection in the total space of the bundle allows to construct the 3-form (\ref{omega-7}) in $\M\times \R$, and the metric arises as the 6D restriction of the 7D metric that is naturally associated with (\ref{omega-7}). 

\subsection{Non-integrable almost complex structures from 7D}

As we have already mentioned, the construction in \cite{BS} gives a $G_2$ holonomy metric in the total space of the $\C^2$ bundle over a 3-dimensional base. Schematically, and using our notations, this construction is as follows. We start with the following 3-form in 7D 
\be\label{omega-7d}
\Omega/2 = -\frac{1}{3} f^3 {\rm Tr}(\E^3) + fg^2 \left( dt{\wedge \rm Tr}\left( \E\W\right) + {\rm Tr}(\E\W^2) \right),
\ee
where $f,g$ are functions of "time" $t$. Requiring this 3-form to be closed gives two simple ODE's on functions $f,g$ whose explicit form we won't need here. The dual 4-form is easily computed by writing the above 3-form in the canonical form. We get
\be\label{omega-7d*}
(\Omega/2)^*/2 = g dt\wedge \left(  -\frac{g^3}{3} {\rm Tr}(\W^3) + g f^2 {\rm Tr}(\E^2\W) \right) - f^2 g^2 {\rm Tr}(\E^2\W^2).
\ee
Requiring this 4-form to be closed gives one more ODE, which is satisfied when the previous two ODE's are satisfied. This makes the associated 7D metric that of $G_2$ holonomy. 

The restriction of (\ref{omega-7d}) to the 6D slice gives a closed 3-form that can be written as the imaginary part of the holomorphic 3-form that is the cube of the Lie algebra valued 1-form $A = g\W+\im f \E$. The real part of this holomorphic 3-form is what appears in brackets in the first term in (\ref{omega-7d*}). This 3-form is not closed, and the closure of $\Omega^*$ does not require it to be closed. Indeed, the equation $d\Omega^*$ just requires that the exterior derivative of the 3-form in brackets in (\ref{omega-7d*}) is proportional to ${\rm Tr}(\E^2\W^2)$. Thus, the associated 6D almost complex structure is not integrable. 

We present this example just to illustrate the point that non-integrable 6D almost complex structures arise naturally from  embedding to 7D. This can be viewed as an argument for why our second proposed interpretation of the dimensional reduction from 6D to 3D is not unnatural. Indeed, recall from the previous section that in the second interpretation one considers the Hitchin function on the space of closed 3-forms parametrised as (\ref{omega**}). For non-zero $\rho$ that signals a non-trivial configuration of the two-form field on the base the arising Euler-Lagrange equations do not imply $d\hat{\Omega}=0$. Instead, one gets $d\hat{\Omega}\sim {\rm Tr}(\E^2\W^2)$, which is similar to what happens in the 7D example we just considered. 

\section{Discussion}

There are two main results of this paper. First, we have provided a 6D interpretation to 3D gravity on $M$ by linking its pure connection formulation to the Hitchin theory of the 3-form $CS(\W)$ on the total space of the ${\rm SU}(2)$ bundle over $M$. Second, we explained how the $\Omega=CS(\W)$ ansatz can be understood as arising via the procedure of the dimensional reduction when the 6D manifold has the structure of the total space $\M$ of the ${\rm SU}(2)$ principal bundle. Thus, we have seen that in general a closed ${\rm SU}(2)$ invariant 3-form on $\M$ can be parametrised by a connection and a 2-form field on the base $M$. When the 2-form field is absent one gets 3D gravity in its pure connection formulation. More generally, one gets gravity coupled to the topological 2-form field. 

In this paper we considered almost exclusively the case of all plus signature with $\Lambda<0$. It is important to remark that all our constructions generalise to other signature $/$ sign of the cosmological constant combinations, with appropriate modifications. Thus, the case of Lorentzian signature, $\Lambda>0$ is obtained from the dimensional reduction with again 3-forms of negative type, and with the 6D manifold fibered with ${\rm SL}(2,\R)$ fibers. The other two cases can be obtained with 3-forms of positive type. In this case, there is no almost complex structure and the arising 6D manifold is not complex. However, the 3-form still defines an operator that acts on the tangent space, and that in this case squares to plus identity. This is what mathematicians refer to as a paracomplex structure. Once the 6D manifold is fibered, this operator can be used to define a connection and a metric. All our constructions then apply, with appropriate sign changes. In particular, the case of 6D manifold with ${\rm SU}(2)$ fibers and with an invariant 3-form of positive type describes gravity with $\Lambda>0$ and all plus signature. When the fibers are copies of ${\rm SL}(2,\R)$ we get $\Lambda<0$ Lorentizan signature gravity. 

We conclude this paper with further remarks about the relation between the 3D and 6D theories. First, the 6D Hitchin theory is topological in the sense that its one-loop partition function, computed in \cite{Pestun:2005rp}, is a ratio of holomoprhic Ray-Singer torsions, see formula (2.65) of this reference. Thus, its dimensional reduction should also be a topological theory, which is what our analysis confirms. The one-loop partition function of 3D gravity is also known, and is also given by an appropriate Ray-Singer torsion. It would be interesting to understand the relation between these two results arising from the dimensional reduction interpretation of this paper. 

Second, the paper \cite{Dijkgraaf:2004te} notices that, at least semi-classically, the partition function of the Hitchin theory exhibits "holomorphic factorisation", see formula (5.30) of this reference and the discussion that follows. In the simplest setup with $\Omega=CS(\W)$, we have interpreted the value of the Hitchin functional as the product of the volume of the gauge group times the value of the 3D gravity action. But the gravity partition function, with the gravity action being the imaginary part of the Chern-Simons functional, see (\ref{grav-cs}), also exhibits the phenomenon of holomorphic factorisation. This factorisation is not as well understood as it deserves to be owing to the difficulties in quantising the ${\rm SL}(2,\C)$ Chern-Simons theory. But it is expected to hold, see \cite{Krasnov:2001cu} for an attempt at understanding this in the setup of asymptotically hyperbolic manifolds. It would be very interesting to relate these two "holomorphic factorisations", as this is likely to shed new light on both 6D and 3D quantum theories. 

The paper \cite{Dijkgraaf:2004te} also proposes to interpret the 6D Hitchin theory partition function as a state is some "topological M theory" in 7D. We have seen in Section \ref{sec:7d} that adding an extra time coordinate is helpful to understand the 6D story. The embedding of 3D gravity into 6D Hitchin theory described in this paper is also relevant to this point, because 3D gravity is two copies of CS theory, and there is a similar phenomenon for the 3D Chern-Simons. Indeed, the value of the Chern-Simons functional $\int_M CS({\bf a})$ on some 3-manifold $M$ can be understood as the value of the integral of the Pontryagin 4-form ${\rm Tr}(\f\wedge\f)$ over some 4-manifold $X$ whose boundary $\partial X=M$ is $M$. There is then a 4D topological field theory \cite{Crane:1994ji} in which the Chern-Simons partition function on $M$ can be interpreted as a state. This seems to be the low-dimensional analog of the 6D to 7D relation envisaged in \cite{Dijkgraaf:2004te}. Again, it would be interesting to determine a relation between these two constructions as comes from the dimensional reduction interpretation of 3D gravity developed in this paper.

Our final remark is that Riemannian signature 3D gravity, at least for the case $\Lambda>0$, is reasonably well understood as a quantum theory. As we have already mentioned in Section \ref{sec:3d}, the quantum theory can be constructed using the Turaev-Viro state sum model \cite{Reshetikhin:1991tc}. The gravity partition function is then the square \cite{Roberts} of the Chern-Simons one, as expected from the action considerations. There is no such good understanding of the more difficult $\Lambda<0$ case as of yet, but work on this is in progress by several research groups. Our interpretation of 3D gravity as sitting inside the 6D Hitchin theory suggests that it should also be possible to construct the 6D quantum theory. It is likely that the case of 3-forms of positive type, which is related to 3D gravity with $\Lambda>0$, should be the simplest starting point. It would be very interesting to attempt to define the quantum theory by some state sum model in 6D, so that this reduces to the Turaev-Viro model when the 6D manifold is of the product form $\M={\rm SU}(2)\times M$. This does not sound impossible, because the sum over spins in the Turaev-Viro model can arise from the expansion of fields on $\M$ into appropriate Fourier modes on ${\rm SU}(2)$. In turn, the 3D understanding may help to construct the 6D quantum theory. It would be very interesting to develop this, but we have nothing else to say on this point at present. 

\section*{Acknowledgement} 

KK and CS were supported by ERC Starting Grant 277570-DIGT\@. 
YH was supported by a grant from ENS Lyon. The authors are grateful to Joel Fine and Yuri Shtanov for discussions on the subject of this paper.

\section*{Appendix}

In this Appendix we carry our the direct computation that starts with parametrisation (\ref{A-twisted}) and confirm the results obtained in the main text by a different method.

\subsection*{Solving the equations}

We can solve equations (\ref{eqs}) as follows. First, let us divide the first equation in the first line by $\alpha^2\delta$ and introduce
\be
x:=\frac{\gamma}{\alpha}, \quad y := \frac{\beta}{\delta}.
\ee
We get
\be\label{y-x}
-y(1-x^2) +2x=0, \quad \Rightarrow \quad y = \frac{2x}{1-x^2}.
\ee
Similarly, we divide the first equation in the second line of (\ref{eqs}) by $\alpha\delta^2$ to get
\be
-y^2 +1 + 2yx=\frac{1}{\alpha\delta^2}.
\ee
Substituting (\ref{y-x}) to the left-hand-side we get
\be\label{d-a}
\frac{1}{\alpha^3 (\delta/\alpha)^2} = \frac{(1+x^2)(1-3x^2)}{(1-x^2)^2}.
\ee
Finally, dividing the second equation in the first line in (\ref{eqs}) by $\alpha^3$ we get
\be\label{a}
\frac{1}{\alpha^3}=1-3x^2.
\ee
Combining this with (\ref{d-a}) we get
\be
\left(\frac{\alpha}{\delta}\right)^2=\frac{1+x^2}{(1-x^2)^2}.
\ee
It remains to determine $x$ in terms of $\rho$ from the last unused equation. Dividing the last equation in the second line of (\ref{eqs}) by $\delta^3$ we get
\be
y^3-3y=\frac{2\rho}{\delta^3}.
\ee
Using the above results for $y,\delta$ in terms of $x$, after some algebra, we get
\be\label{rho-x}
\rho = \frac{x(x^2-3)}{(1+x^2)^{3/2}}.
\ee
This gives the complete solution for $\alpha,\beta,\gamma,\delta$ in terms of $\rho$. In particular, we see that $x=0$ when $\rho=0$.

\subsection*{Calculation of $d\hat{\Omega}$}

In the Section \ref{sec:lift} the condition $d\hat{\Omega}$ together with $d\Omega=0$ was stated as $d\Omega^c=0$, and this in turn was seen to imply that the connection $A$ is flat. It would, however, be wrong to jump to this conclusion in the present case. Indeed, the equation one gets is ${\rm Tr}(dA\wedge A\wedge A)=0$. It can be projected on different vertical-horizontal components, from which the independent equations contained can be extracted. In the case $A=\W+\im \E$ these different independent components all imply $\F=0$. But this is no longer true for the more complicated connection (\ref{A-twisted}). Therefore, there is no shortcut and one needs to compute ${\rm Im}[{\rm Tr}(d A\wedge A\wedge A)]$. We have
\be
AA = (\alpha^2-\gamma^2)\W^2 + (\beta^2-\delta^2)\E^2 +(\alpha\beta-\gamma\delta)(\W\E+\E\W) \\ \nonumber
+ \im \left( 2\alpha\gamma \W^2 +2\beta\delta \E^2+(\alpha\delta+\beta\gamma)(\W\E+\E\W)\right),
\ee
and so
\be\label{d-im}
{\rm Im}[{\rm Tr}(d A\wedge A\wedge A)] = {\rm Tr}\Big( d(\gamma\W+\delta\E) \left[(\alpha^2-\gamma^2)\W^2 + (\beta^2-\delta^2)\E^2 +(\alpha\beta-\gamma\delta)(\W\E+\E\W)\right] \\ \nonumber
+ d(\alpha\W+\beta\E)  \left[ 2\alpha\gamma \W^2 +2\beta\delta \E^2+(\alpha\delta+\beta\gamma)(\W\E+\E\W)\right]\Big).
\ee
We then replace $d\W=\F-\W\W=\E\E-\W\W$, as well as $d\E=g^{-1} \nabla \e g - \W\E-\E\W\equiv \nabla\E -\W\E-\E\W$, where we introduced the notation $g^{-1} \nabla \e g := \nabla\E$, and project on different horizontal-vertical components. We get the following set of equations. First, it is easy to extract the equation that is the three vertical one horizontal component. It is the component proportional to ${\rm Tr}(\W^3)$. This gives
\be\label{w3-eqn}
(\alpha^2-\gamma^2) d\gamma + 2\alpha\gamma d\alpha=0.
\ee
This can be rewritten as
\be
d(\gamma\alpha^2 - \frac{1}{3}\gamma^3) =0.
\ee
Substituting $\alpha,\gamma$ in their parametrisation by $x$ we get
\be
d\left( \frac{x(x^2-3)}{1-3x^2}\right)=0 \quad \Rightarrow \quad dx=0 \quad \Rightarrow \quad d\rho=0.
\ee
This equation is as expected as it says that the field strength of the 2-form field is constant. 

Let us now extract the two vertical two horizontal component of (\ref{d-im}), taking into account the already established fact that $\alpha,\beta,\gamma,\delta$ are constants. We get
\begin{align}
&0=  {\rm Tr}(\W^2\E^2) \left( \gamma(\alpha^2-\gamma^2)-\gamma(\beta^2-\delta^2) +2\alpha(\alpha\gamma -\beta\delta) \right) \\ \nonumber
&-\left(\delta(\alpha\beta-\gamma\delta) +\beta (\alpha\delta+\beta\gamma)\right)  {\rm Tr}((\W\E+\E\W)^2) 
+\left( \delta(\alpha^2-\gamma^2)+2\alpha\beta\gamma\right){\rm Tr}(\nabla\E\W^2) .
\end{align}
We then use
\be
{\rm Tr}((\W\E+\E\W)^2) = - 2{\rm Tr}(\W^2\E^2)
\ee
to simplify the above to
\be
0= {\rm Tr}(\W^2\E^2) \left( \gamma(\alpha^2-\gamma^2)-\gamma(\beta^2-\delta^2)+2\alpha(\alpha\gamma -\beta\delta)  +2\delta(\alpha\beta-\gamma\delta) +2\beta (\alpha\delta+\beta\gamma)\right) \\ \nonumber
+\left( \delta(\alpha^2-\gamma^2)+2\alpha\beta\gamma\right){\rm Tr}(\nabla\E\W^2).
\ee
Substituting the solution for $\alpha,\beta,\gamma,\delta$ that we found in the previous subsection, we find that this equation implies 
\be\label{app-nabla-e}
\nabla \E + \frac{2x(3-x^2)}{(1+x^2)^{3/2}}\E^2=0.
\ee

The last component of (\ref{d-im}) is one vertical three horizontal. The only surviving contribution here is 
\be
0=\left(\delta(\alpha\beta-\gamma\delta) +\beta (\alpha\delta+\beta\gamma)\right) {\rm Tr}(\nabla\E (\E\W+\W\E)).
\ee
However, in view of (\ref{E-eqn}), this equation gives nothing new because ${\rm Tr}(\E^3\W)=0$. 

Thus, to summarise, the whole content of the equation $d\hat{\Omega}=0$ is the statement that $\rho=const$, as well as the equation (\ref{E-eqn}) on $\E$. 

\subsection*{The complex structure}

To compute the geometric connection defined by $J$ we need to write an explicit expression for the complex structure. As in the main text, let us use the dual basis of vector fields defined by (\ref{dual-basis}). Then the vector fields dual to $\alpha\W+\beta\E$ and $\gamma\W + \delta\E$ are given by
\be
\frac{1}{\alpha\delta-\beta\gamma}\left( \delta\frac{\partial}{\partial \W} - \gamma\frac{\partial}{\partial \E}, - \beta \frac{\partial}{\partial \W} + \alpha\frac{\partial}{\partial \E} \right).
\ee
Given that the forms $A$ in (\ref{A-twisted}) give eigenforms of $J_\Omega$ we can write the complex structure operator as follows
\be
J_\Omega = \frac{1}{\im} \left( \alpha \W + \beta \E + \im ( \gamma \W + \delta \E )\right) \frac{1}{2(\alpha\delta-\beta\gamma)}\left( \delta\frac{\partial}{\partial \W} - \gamma\frac{\partial}{\partial \E} - \im\left(- \beta \frac{\partial}{\partial \W} + \alpha\frac{\partial}{\partial \E}  \right)\right) \\ \nonumber
-\frac{1}{\im} \left( \alpha \W + \beta \E - \im ( \gamma \W + \delta \E )\right) \frac{1}{2(\alpha\delta-\beta\gamma)}\left( \delta\frac{\partial}{\partial \W} - \gamma\frac{\partial}{\partial \E} + \im\left(- \beta \frac{\partial}{\partial \W} + \alpha\frac{\partial}{\partial \E}  \right)\right) 
\ee
This gives
\be
J_\Omega = \frac{1}{\alpha\delta-\beta\gamma} ( \gamma \W + \delta \E)\left( \delta\frac{\partial}{\partial \W} - \gamma\frac{\partial}{\partial \E}\right) - \frac{1}{\alpha\delta-\beta\gamma} \left( \alpha \W + \beta \E\right) \left( - \beta \frac{\partial}{\partial \W} + \alpha\frac{\partial}{\partial \E}  \right).
\ee

\subsection*{Horizontal vector fields}

The image under $J_\Omega$ of vertical vector fields is defined to be horizontal. The vertical vector fields are those of the form $\partial/\partial \W\equiv \partial /\partial \m$. We have
\be
J_\Omega \circ \frac{\partial}{\partial \m} = \frac{\alpha\beta+\gamma\delta}{\alpha\delta-\beta\gamma} \frac{\partial}{\partial \W} - \frac{\alpha^2+\gamma^2}{\alpha\delta-\beta\gamma} \frac{\partial}{\partial \E}.
\ee
This is sufficient to define the horizontal lift of vector fields on the base. It is easy to see that the sought lift is given by
\be\label{hor-lift}
\xi_H = \xi_H^\w - \frac{\alpha\beta+\gamma\delta}{\alpha^2+\gamma^2} \, i_\xi \e_\f^i  \frac{\partial}{\partial m^i} = - \frac{\alpha\delta-\beta\gamma}{\alpha^2+\gamma^2} \xi^i \left(  - \frac{\alpha^2+\gamma^2}{\alpha\delta-\beta\gamma} \frac{\partial}{\partial \E^i}+\frac{\alpha\beta+\gamma\delta}{\alpha\delta-\beta\gamma} \frac{\partial}{\partial \W^i} \right),
\ee
where $\xi_H^\w$ is the horizontal lift using the connection $\w$ and $\xi^i \equiv i_\xi \E^i$.

\subsection*{The metric}

To compute the metric, we take the horizontal lifts of two basic vector fields, apply to the lifts the complex structure and then take the metric contraction using the metric in the fiber. We have
\be
J_\Omega \circ \xi_H = - \frac{\alpha\delta-\beta\gamma}{\alpha^2+\gamma^2} \xi^i J_\Omega \circ J_\Omega \circ \frac{\partial}{\partial m^i} =  \frac{\alpha\delta-\beta\gamma}{\alpha^2+\gamma^2} \xi^i \frac{\partial}{\partial m^i}.
\ee
This shows that the metric as defined by the complex structure together with the metric in the fibers is given by
\be
(\xi,\eta)_J = \left( \frac{\alpha\delta-\beta\gamma}{\alpha^2+\gamma^2} \right)^2 \xi^i \eta^j \delta_{ij}.
\ee
In other words, the metric is as corresponds to the frame field
\be\label{E-J}
\E_{J}:= \frac{\alpha\delta-\beta\gamma}{\alpha^2+\gamma^2}  \E =\frac{1-3x^2}{(1+x^2)^{3/2}} \e=\sqrt{1-\rho^2}\, \E.
\ee
Here we have substituted the values of $\alpha,\beta,\gamma,\delta$ in their parametrisation by $x$, and then expressed the final result in terms of $\rho$ (\ref{rho-x}). 

\subsection*{The connection}

The complex structure also defines a new set of connection 1-forms, such that the horizontal vector fields (\ref{hor-lift}) are in its kernel. It is easy to see that the connection 1-forms are given by
\be\label{W-J}
\W_J := \W + \frac{\alpha\beta+\gamma\delta}{\alpha^2+\gamma^2} \E = \W+ \frac{x(3-x^2)}{(1+x^2)^{3/2}}\E = \W-\rho \,\E.
\ee
Indeed, we have $i_{\xi_H} \W_J=0$ as required. To write the last equality we have used (\ref{rho-x}). Thus, we have established by a direct computation that the geometric connection defined by $\Omega$ is $\W_J=\W-\rho\E$. The equation (\ref{app-nabla-e}) is then the statement that this connection is torsion free.

\end{document}